\documentclass[aps,prd,twocolumn,10pt]{revtex4-1}
\usepackage{amssymb}
\usepackage{graphicx}
\usepackage{amsmath}
\usepackage{hyperref}
\usepackage{subfigure}
\usepackage{multirow}
\usepackage{setspace}
\usepackage{verbatim}
\usepackage{tensor}
\usepackage{float}
\usepackage{color}
\usepackage{ulem}
\usepackage[utf8]{inputenc}

\begin{document}

\title{Cosmic inflation from broken conformal symmetry}

\author{Rong-Gen Cai$^{1,2,3}$}
\email{cairg@itp.ac.cn}

\author{Yu-Shi Hao$^{1,3}$}
\email{haoyushi@itp.ac.cn}

\author{Shao-Jiang Wang$^{1}$}
\email{schwang@itp.ac.cn (corresponding author)}

\affiliation{$^1$CAS Key Laboratory of Theoretical Physics, Institute of Theoretical Physics, Chinese Academy of Sciences, Beijing 100190, China}
\affiliation{$^2$School of Fundamental Physics and Mathematical Sciences, Hangzhou Institute for Advanced Study (HIAS), University of Chinese Academy of Sciences (UCAS), Hangzhou 310024, China}
\affiliation{$^3$School of Physical Sciences, University of Chinese Academy of Sciences (UCAS), Beijing 100049, China}

\begin{abstract}
A period of rapidly accelerating expansion is expected in the early Universe implemented by a scalar field slowly rolling down along an asymptotically flat potential preferred by the current data. In this paper, we point out that this picture of the cosmic inflation with an asymptotically flat potential could emerge from the Palatini quadratic gravity by adding the matter field in such a way to break the local gauged conformal symmetry in both kinetic and potential terms. 
\end{abstract}
\maketitle

\section{Introduction} \label{sec:introduction}

A simultaneous resolution for the fine-tuned horizon problem, flatness problem, and monopole problem calls for a period of rapidly accelerating expansion of spacetime~\cite{Brout:1977ix,Starobinsky:1980te,Kazanas:1980tx,Sato:1980yn,Guth:1980zm,Linde:1981mu,Albrecht:1982wi,Linde:1983gd} in the early Universe at least prior to the big bang nucleosynthesis. This inflationary paradigm also provides the causal productions for the primordial cosmological perturbations with a nearly scale-invariant spectrum \cite{Mukhanov:1981xt,Mukhanov:1982nu,Hawking:1982cz,Guth:1982ec,Starobinsky:1982ee,Bardeen:1983qw,Kodama:1984ziu,Mukhanov:1985rz} responsible for the observed cosmic microwave background \cite{WMAP:2012nax,Planck:2018jri} and large scale structures \cite{BOSS:2016wmc,eBOSS:2020yzd}. The standard realization for such an inflationary period usually turns to a slow-roll scalar field along with some inflationary potential \cite{Linde:1983gd}. The most recent constraint \cite{BICEPKeck:2021gln} on the cosmic inflation still prefers a single-field slow-roll plateau-like potential. 

There are two popular implements for such a plateau-like potential: the simplest one is the Starobinsky inflation \cite{Starobinsky:1980te} with additional quadratic term for the Ricci scalar curvature $R$;  the most economic one is the Higgs inflation \cite{Bezrukov:2007ep} with the only known fundamental scalar field (Higgs boson) so far as the inflaton non-minimally coupled to $R$. It was realized in recent years that they could be all constructed in general from the cosmological attractors \cite{Galante:2014ifa} to consist of the $\alpha$-attractors \cite{Kallosh:2013hoa,Kallosh:2013daa,Ferrara:2013rsa,Kallosh:2013yoa,Linde:2015uga} (including the Starobinsky inflation as a special case  \cite{Cai:2014bda}) and $\xi$-attractors \cite{Kallosh:2013tua} (including the Higgs inflation and induced inflation \cite{Giudice:2014toa,Pallis:2013yda,Pallis:2014dma,Pallis:2014boa,Kallosh:2014rha} as special cases).  

It is then intriguing to explore the theoretical origin of these asymptotically flat potentials. The current observational data merely reveals two clues: (i) A plateau-like potential is supposed to admit an approximate shift symmetry, which should be slightly broken to protect an asymptotically flat potential against quantum corrections. (ii) A nearly scale-invariant spectrum of primordial perturbations suggests a slightly broken scale symmetry in the very early Universe from de Sitter (dS) to quasi-dS phases. An appealing understanding of the cosmic inflation should explain  the roles played by these two symmetries.

Motivated by the superconformal approach \cite{Kallosh:2000ve,Ferrara:2010yw,Ferrara:2010in} to the Higgs-like inflation and Starobinsky inflation \cite{Kallosh:2013pby,Kallosh:2013lkr}, the $\alpha$-attractor approach is able to really appreciate the role played by the  conformal (scale) symmetry. The starting point of this approach is an old observation that a single real conformal compensator (a scalar field called conformon) with the Lagrangian $\sqrt{-g}(\frac1{12}\varphi^2R+\frac12(\partial\varphi)^2-\frac14\lambda\varphi^4)$ is equivalent to the pure Einstein gravity with a positive cosmological constant  $9\lambda$ (thus a dS solution) after gauge-fixing the conformon field to some constant thanks to the local conformal symmetry of the Lagrangian. 

Although the gauge-fixing for the conformon field eliminates the concern for the presence of ghost from the wrong-sign kinetic term, the conformon field cannot be gauge-fixed if one tries to construct any nontrivial structure (namely inflation with quasi-dS phase) by explicitly breaking the local conformal symmetry. Therefore, the $\alpha$-attractor approach introduces an extra scalar field with a joint global symmetry \cite{Kallosh:2013pby,Kallosh:2013lkr,Kallosh:2013hoa,Kallosh:2013daa} with the conformon field but still leaves the local conformal symmetry unbroken in order to fix the gauge of the would-be-ghost conformon field. After gauge-fixing, the local conformal symmetry is spontaneously broken, and the global-symmetry-breaking potential leads to an asymptotically flat potential. However, the global symmetry for a successful inflationary implementation is restricted due to  the wrong-sign kinetic term required by the local conformal symmetry.

The introduction of the conformon field with wrong-sign kinetic term could be avoided if one dives into the Palatini formalism of gravity \cite{Hehl:1994ue,Gronwald:1997bx} where the metric and affine connection are treated as independent degrees of freedom. In the Palatini formalism, the conformon field with wrong-sign kinetic term naturally emerges as a geometric gauge degree of freedom from the $R^2$ term (see Eq.~\eqref{eq:SOSR2} below), which has been already derived but overlooked in \cite{Ghilencea:2020piz}. The focus there is mainly on the  dynamical recovering of the metric Einstein gravity in the absence of matter field in the Palatini formalism of a general quadratic gravity with the local conformal symmetry. The metric Einstein gravity therefore emerges at the decoupling limit of the Weyl gauge field after eating up the dilaton field $\partial_\mu\ln\varphi^2$ with a shift symmetry inherited from the local gauged conformal symmetry of $\varphi$. See \cite{Ghilencea:2018dqd,Ghilencea:2019jux,Ghilencea:2019rqj} for a similar realization in the Weyl quadratic gravity and a comparison to the Palatini quadratic gravity \cite{Ghilencea:2020rxc} as well as its concrete realizations in the standard model of particle physics \cite{Ghilencea:2021lpa} and cosmology \cite{Ghilencea:2021jjl}. See also \cite{Tang:2018mhn,Tang:2019uex,Tang:2019olx,Tang:2020ovf,Tang:2021lcn} \cite{SravanKumar:2018tgk} for other trials.

However, to carry out  an inflationary potential in the Palatini formalism in a conformally invariant manner, it seems that a global symmetry shared with an additional scalar field is still needed to be slightly broken \cite{Mikura:2020qhc,Mikura:2021ldx} similar to the $\alpha$-attractor approach. Nevertheless, we will point out in this paper that, in the Palatini quadratic gravity, the presence of an additional global symmetry is not necessary as also expected from the swampland conjecture \cite{Banks:1988yz,Banks:2010zn,Harlow:2018tng,Harlow:2018jwu} of no global symmetry in quantum gravity. Without introducing any global symmetry, a plateau-like inflationary potential is always implied when the matter field is included in such a way to appropriately break the local conformal symmetry.

The outline of this paper is as follows: In Sec. \ref{sec:quadratic}, we review  previous the results on the emergence of metric Einstein gravity from Palatini quadratic gravity. In Sec \ref{sec:addmatter}, we show the emergence of non-plateau-like and plateau-like inflation models when adding the matter field differently in terms of the local conformal symmetry. We summarize our results and discuss possible future perspectives in Sec. \ref{sec:conclusion}.
The convention for metric $g_{\mu\nu}$ is $(-,+,+,+)$, the Planck mass is $M_\mathrm{Pl}\equiv1/\sqrt{8\pi G_N}$, and quantities with an overbar symbol (like the Ricci scalar $\bar{R}$ and covariant derivative $\bar{\nabla}$) are always subjected to the Levi-Civita connection $\bar{\Gamma}^\rho_{\mu\nu}=\frac12g^{\rho\lambda}(\partial_\mu g_{\nu\lambda}+\partial_\nu g_{\lambda\mu}-\partial_\lambda g_{\mu\nu})$. The Riemann tensor and its variation under the connection variation $\Gamma^\rho_{\mu\nu}\to\Gamma^\rho_{\mu\nu}+\delta\Gamma^\rho_{\mu\nu}$ read $R^\rho_{\,\,\mu\sigma\nu}=\partial_\sigma\Gamma^\rho_{\mu\nu}-\partial_\nu\Gamma^\rho_{\mu\sigma}+\Gamma^\rho_{\lambda\sigma}\Gamma^\lambda_{\mu\nu}-\Gamma^\rho_{\lambda\nu}\Gamma^\lambda_{\mu\sigma}$ and $\delta R^\rho_{\,\,\mu\sigma\nu}=\nabla_\sigma(\delta\Gamma^\rho_{\mu\nu})-\nabla_\nu(\delta\Gamma^\rho_{\mu\sigma})+T^\lambda_{\,\,\sigma\nu}\delta\Gamma^\rho_{\mu\lambda}$, respectively, where the torsion tensor $T^\rho_{\mu\nu}=\Gamma^\rho_{\mu\nu}-\Gamma^\rho_{\nu\mu}$ will be simply set to zero hereafter for convenience due to the geometric trinity of gravity \cite{Jimenez:2019woj}. 
We remind here that the geometric trinity of gravity is an equivalence among three different ways to describe gravity: the traditional way of using Riemann tensor $R^\alpha_{\,\,\beta\mu\nu}$ in general relativity describes a rotation of vector after transported in parallel along a closed curve, while the torsion tensor $T^\rho_{\mu\nu}=\Gamma^\rho_{\mu\nu}-\Gamma^\rho_{\nu\mu}$  in teleparallel equivalent of general relativity describes the non-closure of parallelograms formed by two vectors transported along each other, and the non-metricity tensor $Q_\lambda^{\,\,\mu\nu}=\nabla_\lambda g^{\mu\nu}$ in symmetric teleparallel equivalent of general relativity describes the dilation of the length of a vector when transported along a curve. This geometric trinity of gravity might be jeopardized when the matter field is added. We therefore leave the case with the presence of the torsion field for future work.

\section{Palatini quadratic gravity}\label{sec:quadratic}

In this section, we review the Palatini quadratic gravity with a local conformal symmetry, which reduces to the metric Einstein gravity with a positive cosmological constant when fixing the gauge of the local conformal symmetry. Although most of the derivations in this section have been presented before in \cite{Ghilencea:2020piz}, we re-derive these results to set up our notations and conventions to be used later on.

\subsection{Palatini $R^2$ gravity}\label{subsec:R2}

We start with the Palatini $R^2$ gravity with an action of a form
\begin{align}\label{eq:SR2}
S[g,\Gamma]=\int\mathrm{d}^4x\sqrt{-g}\,\,\frac{\alpha}{2}R(g,\Gamma)^2, \quad \alpha>0,
\end{align}
whose field equation for metric reads
\begin{align}
R\left(R_{(\mu\nu)}-\frac{R}{4}g_{\mu\nu}\right)=0.
\end{align}
Despite of the trivial solution $R=0$, the non-trivial part of the field equation for metric is (differ from the usual GR case with an extra factor of $1/2$ in front of the Ricci scalar)
\begin{align}\label{eq:EOMg}
R_{(\mu\nu)}-\frac{R}{4}g_{\mu\nu}=0,
\end{align}
whose trace is identically satisfied (unlike the usual GR case that the trace part of the Einstein field equation gives rise to the vacuum solution $R=0$). Therefore, despite of the trivial solution $R=0$, the trace part of field equation for metric puts no constraint on Ricci scalar $R$, and the only constraint on $R$ comes from the equation of motion (EoM) for the symmetric connection,
\begin{align}
\nabla_\lambda\left(\sqrt{-g}Rg^{\mu\nu}\right)-\nabla_\rho\left(\sqrt{-g}Rg^{\rho(\mu}\right)\delta_\lambda^{\nu)}=0,
\end{align}
which, after substituted with its trace in $\lambda=\nu$ by $\nabla_\rho\left(\sqrt{-g}Rg^{\rho\mu}\right)=0$, becomes
\begin{align}
\nabla_\lambda\left(\sqrt{-g}Rg^{\mu\nu}\right)=0.
\end{align}
Expanding above equation as
\begin{align}
g^{\mu\nu}\partial_\lambda R-\frac{R}{2}g^{\mu\nu}Q_\lambda+RQ_\lambda^{\,\,\,\mu\nu}=0
\end{align}
by the non-metricity tensor $Q_\lambda^{\,\,\,\mu\nu}=\nabla_\lambda g^{\mu\nu}$ and non-metricity vector $Q_\lambda=g_{\mu\nu}\nabla_\lambda g^{\mu\nu}$ followed by contracted with $g_{\mu\nu}$, one has
\begin{align}\label{eq:EOMGamma}
\partial_\lambda(\ln R)=\frac14Q_\lambda.
\end{align}
Note that the action \eqref{eq:SR2} is actually a redundant description due to the local conformal symmetry, $S[g,\Gamma]=S[\tilde{g}, \tilde{\Gamma}]$, under the local conformal transformations,
\begin{align}\label{eq:CT}
\tilde{g}_{\mu\nu}=\Omega(x)^2g_{\mu\nu},\quad \tilde{\Gamma}^\rho_{\mu\nu}=\Gamma^\rho_{\mu\nu},
\end{align}
since the Ricci scalar-square $R(g,\Gamma)^2=(g^{\mu\nu}R_{\mu\nu}(\Gamma))^2=(\Omega^2\tilde{g}^{\mu\nu}R_{\mu\nu}(\tilde{\Gamma}))^2\equiv\Omega^4\tilde{R}^2$  compensates the contribution from $\sqrt{-g}=\Omega^{-4}\sqrt{-\tilde{g}}$. To fix  this gauge symmetry, one should fix one of the scalar degree of freedom, for example, gauge-fixing $R$ to some constant $C\neq0$. Then, the Eq. \eqref{eq:EOMGamma} reduces to the vanishing non-metricity with the metric compatible Levi-Civita connection, and the Eq. \eqref{eq:EOMg} reduces to the usual GR case of the Einstein field equation with a non-vanishing cosmological constant $\Lambda=C/4$. Note that the vacuum solution $R=0$ automatically satisfies the connection EOM \eqref{eq:EOMGamma}, therefore, only in this case, it does not reduce to the metric Einstein gravity with a cosmological constant. In what follows, we will not consider the case with $R=0$.

We can also introduce an auxiliary field $\varphi^2/2=F'(\phi)=\alpha\phi$ in the expansion of  $F(R)=F(\phi)+F'(\phi)(R-\phi)$ for $F(R)=(\alpha/2)R^2$, and then arrive at an equivalent Jordan-frame action
\begin{align}\label{eq:SR2varphi}
S[g,\Gamma;\varphi]=\int\mathrm{d}^4x\sqrt{-g}\left(\frac{\varphi^2}{2}R(g,\Gamma)-\frac{\varphi^4}{8\alpha}\right),
\end{align}
which reduces to \eqref{eq:SR2} when putting $\varphi$-field on-shell  by its EoM $\varphi^2/2=\alpha R$. This Jordan-frame action enjoys a local gauged conformal symmetry, $S[g, \Gamma; \varphi]=S[\tilde{g}, \tilde{\Gamma}; \tilde{\varphi}]$, under the local gauged conformal transformations
\begin{align}\label{eq:GCT1}
\tilde{g}_{\mu\nu}=\Omega^2g_{\mu\nu},\quad\tilde{\Gamma}^\rho_{\mu\nu}=\Gamma^\rho_{\mu\nu},\quad\tilde{\varphi}=\Omega^{-1}\varphi,
\end{align}
where $\varphi$ is actually a gauge degree of freedom of the shift symmetry $\ln\tilde{\varphi}=\ln\varphi-\ln\Omega$ compensating the local conformal transformation \eqref{eq:CT}. However, unlike in the metric formalism, the auxiliary field $\varphi$ is not a dynamical degree of freedom. This could be seen after conformally transforming \eqref{eq:SR2varphi} into the Einstein-frame action as
\begin{align}\label{eq:SER2}
S[\tilde{g},\tilde{\Gamma}]&\equiv S[g_{\mu\nu}=\Omega^{-2}\tilde{g}_{\mu\nu},\Gamma^\rho_{\mu\nu}=\tilde{\Gamma}^\rho_{\mu\nu};\varphi]\nonumber\\
&=\int\mathrm{d}^4x\sqrt{-\tilde{g}}\left(\frac{M_\mathrm{Pl}^2}{2}R(\tilde{g},\tilde{\Gamma})-\frac{M_\mathrm{Pl}^4}{8\alpha}\right)
\end{align}
with a specific conformal factor $\Omega(x)^2=\varphi(x)^2/M_\mathrm{Pl}^2$. Note that $\varphi$ remains unchanged during the local conformal transformations \eqref{eq:CT} and it only transforms as $\tilde{\varphi}=\Omega^{-1}\varphi$ when testing for the local gauged conformal symmetry. It is easy to see that this Einstein-frame action $S[\tilde{g},\tilde{\Gamma}]$ is equivalent to the Jordan-frame action $S[g,\Gamma;\varphi]$ by directly gauge-fixing $\varphi$ to $M_\mathrm{Pl}$ thanks to the local gauged conformal symmetry of $\varphi$. Now that the Einstein-frame action is minimally coupled, putting the connection on-shell reproduces the Levi-Civita connection, and the metric-affine geometry reduces to the Riemannian geometry.  Hence the metric Einstein gravity is recovered in a gauge-fixing form but with an additional positive cosmological constant.

Equivalently, Ref. \cite{Ghilencea:2020piz} provides alternative treatment on the action \eqref{eq:SR2varphi} by first putting the connections on-shell before making either local conformal transformations \eqref{eq:CT} or gauge-fixing $\varphi$ to $M_\mathrm{Pl}$.  Note that the torsionless version of Stokes' theorem in Palatini formalism renders $\int\mathrm{d}^4x\nabla_\mu(\sqrt{-g}V^\mu)=0$, one obtains the EoM of the connection,
\begin{align}\label{eq:EOMGammaR2}
\nabla_\lambda(\sqrt{-g}\varphi^2g^{\mu\nu})-\nabla_\rho(\sqrt{-g}\varphi^2g^{\rho(\mu})\delta^{\nu)}_\lambda=0,
\end{align}
which, after contracting $\nu=\lambda$, gives rise to an equation $\nabla_\nu(\sqrt{-g}\varphi^2g^{\mu\nu})=0$ that could be rewritten as $\nabla_\nu(\sqrt{-f}f^{\mu\nu})=0$ in terms of a metric-compatible auxiliary metric $f_{\mu\nu}\equiv\varphi^2g_{\mu\nu}$. Therefore, the connection could be solved as the Levi-Civita connection $\Gamma^\rho_{\mu\nu}(f)=\frac12f^{\rho\lambda}(\partial_\mu f_{\nu\lambda}+\partial_\nu f_{\lambda\mu}-\partial_\lambda f_{\mu\nu})$ in terms of $f_{\mu\nu}$, which, after expressed in terms of $g_{\mu\nu}$ explicitly, becomes
\begin{align}\label{eq:GammaR2}
\Gamma^\rho_{\mu\nu}=\bar{\Gamma}^\rho_{\mu\nu}(g)+\frac12(G_\mu\delta^\rho_\nu+G_\nu\delta^\rho_\mu-G^\rho g_{\mu\nu}),
\end{align}
with abbreviating $G_\mu\equiv\partial_\mu\ln\varphi^2=\bar{\nabla}_\mu\ln\varphi^2=\nabla_\mu\ln\varphi^2$. Note that with on-shell connection, the Weyl gauge field $A_\mu\equiv\frac12(\Gamma^\rho_{\mu\rho}-\bar{\Gamma}^\rho_{\mu\rho}(g))=G_\mu$ is fixed and determined by $G_\mu$ field alone, which is in fact related to the fact that the action \eqref{eq:SR2varphi} is invariant under the projective transformation $\tilde{\Gamma}^\rho_{\mu\nu}=\Gamma^\rho_{\mu\nu}+\delta^\rho_\mu\xi_\nu(x)$ for an arbitrary vector field $\xi_\mu(x)$ used for gauge-fixing $A_\mu$.  Putting the connection $\Gamma^\rho_{\mu\nu}$ on-shell (OS) with solution \eqref{eq:GammaR2}, the Ricci scalar reads $R(g,\Gamma_\mathrm{OS})=\bar{R}(g)-3\bar{\nabla}_\mu G^\mu-\frac32G_\mu G^\mu$, and the action \eqref{eq:SR2varphi} becomes
\begin{align}\label{eq:SOSR2}
S[g ; \varphi]=\int\mathrm{d}^4x\sqrt{-g}\left(\frac{\varphi^2}{2}\bar{R}(g)+3(\bar{\nabla}_\mu\varphi)^2-\frac{\varphi^4}{8\alpha}\right),
\end{align}
which is exactly the Lagrangian form with a wrong-sign kinetic term desired by the $\alpha$-attractor approach in the first place. The on-shell action \eqref{eq:SOSR2} also enjoys a local gauged conformal symmetry, $S[g;\varphi]=S[\tilde{g};\tilde{\varphi}]$, under the local gauged conformal transformations
\begin{align}\label{eq:GCT2}
\tilde{g}_{\mu\nu}=\Omega^2g_{\mu\nu}, \quad \tilde{\varphi}=\Omega^{-1}\varphi,
\end{align}
thanks to the plus sign of $+3(\bar{\nabla}_\mu\varphi)^2$ (namely conformon) that is crucial for exact cancellations with respect to the $\Omega$-dependent terms in $\bar{R}(g)=\Omega^2[\bar{R}(\tilde{g})+3\tilde{\bar{\nabla}}^2\ln\Omega^2-\frac32(\tilde{\bar{\nabla}}_\mu\ln\Omega^2)^2]$. Now that $\varphi$ is a gauge degree of freedom, one can either directly gauge-fix $\varphi$ to $M_\mathrm{Pl}$ or choose a specific conformal factor $\Omega^2=\varphi^2/M_\mathrm{Pl}^2$ to conformally transform \eqref{eq:SOSR2} via $S[g_{\mu\nu}=\Omega^{-2}\tilde{g}_{\mu\nu}; \varphi]$  as
\begin{align}
S[\tilde{g}]=\int\mathrm{d}^4x\sqrt{-\tilde{g}}\left(\frac{M_\mathrm{Pl}^2}{2}\bar{R}(\tilde{g})-\frac{M_\mathrm{Pl}^4}{8\alpha}\right),
\end{align}
which is exactly the action \eqref{eq:SER2} with on-shell connection.

In a short summary, the $R^2$ term in the Palatini formalism contributes an extra non-dynamical gauge degree of freedom $\varphi$ of shift symmetry $\ln\tilde{\varphi}^2=\ln\varphi^2-\ln\Omega^2$ under the local gauged conformal transformations \eqref{eq:GCT1} or \eqref{eq:GCT2}. Therefore, $\ln\varphi^2$ and $G_\mu=\partial_\mu\ln\varphi^2$ behave like the dilaton field and the would-be Goldstone field, respectively. After gauge-fixing $\varphi$ to $M_\mathrm{Pl}$, the metric Einstein gravity with a positive cosmological constant is recovered.

\subsection{Palatini $R^2+R_{[\mu\nu]}^2$ gravity}\label{subsec:R2F2}

In the Palatini formalism, the Ricci tensor $R_{\mu\nu}$ receives  an anti-symmetric contribution $R_{[\mu\nu]}\equiv\frac12(R_{\mu\nu}-R_{\nu\mu})=\frac12(\partial_\mu\Gamma^\rho_{\rho\nu}-\partial_\nu\Gamma^\rho_{\rho\mu})$. It is easy to show that the difference between the Palatini connection and Levi-Civita connection is transformed as a tensor, then $R_{[\mu\nu]}$ resembles the Maxwell-like field strength tensor,
\begin{align}
R_{[\mu\nu]}=\partial_\mu A_\nu-\partial_\nu A_\mu\equiv F_{\mu\nu},
\end{align}
if one defines the Weyl gauge field $A_\mu=\frac12(\Gamma_\mu-\bar{\Gamma}_\mu(g))$ with abbreviations $\Gamma_\mu\equiv\Gamma^\rho_{\rho\mu}$ and $\bar{\Gamma}_\mu(g)\equiv\bar{\Gamma}^\rho_{\rho\mu}(g)=\partial_\mu\ln\sqrt{-g}$. We therefore turn to the Palatini $R^2+R_{[\mu\nu]}^2$ gravity with an action of form
\begin{align}\label{eq:SR2F2}
S[g, \Gamma]=\int\mathrm{d}^4x\sqrt{-g}\left(\frac{\alpha}{2}R(g,\Gamma)^2-\frac{R_{[\mu\nu]}^2(\Gamma)}{4\beta^2}\right),
\end{align}
which also exhibits the local conformal symmetry, $S[g, \Gamma]=S[\tilde{g}, \tilde{\Gamma}]$, under the local conformal transformation \eqref{eq:CT} since $R_{[\mu\nu]}(\Gamma)^2\equiv R_{[\mu\nu]}(\Gamma)R^{[\mu\nu]}(\Gamma)=\Omega^4\tilde{g}^{\rho\mu}\tilde{g}^{\sigma\nu}R_{[\mu\nu]}(\tilde{\Gamma})R_{[\rho\sigma]}(\tilde{\Gamma})\equiv\Omega^4R_{[\mu\nu]}(\tilde{\Gamma})^2$ compensates the contribution from $\sqrt{-g}=\Omega^{-4}\sqrt{-\tilde{g}}$. Similar to Sec.~\ref{subsec:R2}, one can also introduce an auxiliary scalar $\varphi$ to rewrite $\alpha^2R^2=\varphi^2R-\varphi^4/(4\alpha)$, and then the action becomes
\begin{align}\label{eq:SR2F2varphi}
S[g, \Gamma; \varphi]=\int\mathrm{d}^4x\sqrt{-g}\left(\frac{\varphi^2}{2}R(g,\Gamma)-\frac{F_{\mu\nu}^2(A)}{4\beta^2}-\frac{\varphi^4}{8\alpha}\right),
\end{align}
which also enjoys a local gauged conformal symmetry, $S[g, \Gamma; \varphi]=S[\tilde{g}, \tilde{\Gamma}; \tilde{\varphi}]$, under the local gauged conformal transformations \eqref{eq:GCT1}. Note that $\tilde{A}_\mu=A_\mu-\partial_\mu\ln\Omega^2$ does not transform independently from the local conformal transformations \eqref{eq:CT} but inherited from $\bar{\Gamma}_\mu(g)=\tilde{\bar{\Gamma}}_\mu(\tilde{g})-2\partial_\mu\ln\Omega^2$ under the local conformal transformations \eqref{eq:CT}. It is easy to see that  both \eqref{eq:SR2F2} and \eqref{eq:SR2F2varphi} admit additional gauge shfit symmetry under  $\tilde{A}_\mu=A_\mu-\partial_\mu\omega^2$ for an arbitrary gauge function $\omega(x)$, and hence $A_\mu$ is actually a gauge degree of freedom. It is worth noting that this gauge shift symmetry of $A_\mu$ is different from the gauge shfit symmetry of $\varphi$ since $\omega$ does not need to be coincided with the local conformal transformation factor $\Omega$. 

Alternatively, Ref. \cite{Ghilencea:2020piz}  provides another intriguing view on the action \eqref{eq:SR2F2varphi} by first putting the connection on-shell before making either local conformal transformations \eqref{eq:CT} or gauge-fixing $\varphi$ to $M_\mathrm{Pl}$. The EoM of the connection is obtained as
\begin{align}\label{eq:EOMGammaR2F2}
\nabla_\lambda(\sqrt{-g}\varphi^2g^{\mu\nu})-\nabla_\rho(\sqrt{-g}\varphi^2g^{\rho(\mu})\delta^{\nu)}_\lambda=\frac{\nabla_\rho(\sqrt{-g}F^{\rho(\mu})\delta^{\nu)}_\lambda}{\beta^2},
\end{align}
which, after contracting $\lambda=\nu$, gives rise to an equation $5\nabla_\rho(\sqrt{-g}F^{\mu\rho})=3\beta^2\nabla_\nu(\sqrt{-g}\varphi^2g^{\mu\nu})$. Plugging this equation back to the  EoM of connection leads to $\nabla_\lambda(\sqrt{-g}\varphi^2g^{\mu\nu})=\frac25\nabla_\rho(\sqrt{-g}\varphi^2g^{\rho(\mu})\delta^{\nu)}_\lambda$.
This inspires an ansatz as $\nabla_\lambda(\sqrt{-g}\varphi^2g^{\rho\sigma})=k\sqrt{-g}\varphi^2(V^\rho\delta^\sigma_\lambda+V^\sigma\delta^\rho_\lambda)$ for arbitrary number $k$, which, after multiplying both sides of the ansatz with $g_{\rho\sigma}$ and appreciating $\nabla_\lambda\ln\sqrt{-g}=-\frac12g_{\mu\nu}\nabla_\lambda g^{\mu\nu}=\frac12g^{\mu\nu}\nabla_\lambda g_{\mu\nu}$, could be solved with $kV_\lambda=\nabla_\lambda\ln(\sqrt{-g}\varphi^4)$.
One can also multiply both sides of the ansatz with $g_{\mu\rho}g_{\nu\sigma}$ and then use $\nabla_\lambda\ln\sqrt{-g}=kV_\lambda-\nabla_\lambda\ln\varphi^4$ to obtain $\nabla_\lambda(\varphi^2g_{\mu\nu})=k\varphi^2(g_{\mu\nu}V_\lambda-g_{\mu\lambda}V_\nu-g_{\nu\lambda}V_\mu)$. Now the connection could be solved as
\begin{align}\label{eq:GammaR2F2}
\Gamma^\rho_{\mu\nu}=\bar{\Gamma}^\rho_{\mu\nu}(\varphi^2g)+\frac{k}{2}(3V^\rho g_{\mu\nu}-V_\mu\delta^\rho_\nu-V_\nu\delta^\rho_\mu),
\end{align}
with $\bar{\Gamma}^\rho_{\mu\nu}(\varphi^2g)$ the Levi-Civita connection subjected to an auxiliary metric $f_{\mu\nu}\equiv\varphi^2g_{\mu\nu}$, which could be further expressed in terms of $g_{\mu\nu}$ as $\bar{\Gamma}^\rho_{\mu\nu}(\varphi^2g)=\bar{\Gamma}^\rho_{\mu\nu}(g)+\frac12(G_\mu\delta^\rho_\nu+G_\nu\delta^\rho_\mu-G^\rho g_{\mu\nu})$.
Therefore, the Weyl gauge field $A_\mu=\frac12(\Gamma_\mu-\bar{\Gamma}_\mu(g))$ with on-shell connection could be obtained as
\begin{align}
A_\mu=G_\mu-\frac{k}{2}V_\mu=-\frac12\nabla_\mu\ln\sqrt{-g}=\frac14g_{\rho\sigma}\nabla_\mu g^{\rho\sigma},
\end{align}
which is nothing but a quarter of the non-metricity vector field $Q_\mu\equiv g_{\rho\sigma}\nabla_\mu g^{\rho\sigma}$. The Ricci scalar with on-shell connection is derived as  $R(g, \Gamma_\mathrm{OS})=\bar{R}(g)-3\bar{\nabla}_\mu G^\mu-\frac32G_\mu G^\mu+6k(\bar{\nabla}_\mu V^\mu+G_\mu V^\mu)-\frac32k^2 V_\mu V^\mu$. Using $\varphi^2(\bar{\nabla}_\mu V^\mu+G_\mu V^\mu)=\bar{\nabla}_\mu(\varphi^2V^\mu)$, the action \eqref{eq:SR2F2varphi} with on-shell connection becomes
\begin{align}\label{eq:SOSR2F2}
S[g, A &; \varphi]=\int\mathrm{d}^4x\sqrt{-g}\left(\frac{\varphi^2}{2}\bar{R}(g)-\frac{1}{4\beta^2}F_{\mu\nu}(A)^2\right.\nonumber\\
&\left.+3(\bar{\nabla}_\mu\varphi)^2-3\varphi^2(A_\mu-G_\mu(\varphi))^2-\frac{\varphi^4}{8\alpha}\right).
\end{align}

Note that at this point $A_\mu$ does not enjoy the arbitrary gauge shift symmetry under $\tilde{A}_\mu=A_\mu-\partial_\mu\omega^2$ anymore. It seems that  putting the connection on shell picks out a particular gauge choice $\omega=\Omega$ for $A_\mu$ when transformed coherently with the local gauged conformal transformations \eqref{eq:GCT2}. Note also that, putting the connection on-shell does not fix all its components but leaves $A_\mu$ undetermined since contracting $\rho=\nu$ in \eqref{eq:GammaR2F2} simply reduces to a trivial identity. This is caused by the explicitly broken projective symmetry of \eqref{eq:SR2F2varphi} and \eqref{eq:SOSR2F2} under the projective transformation $\tilde{\Gamma}^\rho_{\mu\nu}=\Gamma^\rho_{\mu\nu}+\delta^\rho_\mu\xi_\nu(x)$ for an arbitrary vector field $\xi_\mu(x)$, which would otherwise fix the Weyl gauge field $A_\mu$. This is different from the case in Sec. \ref{subsec:R2} where $A_\mu$ is fully determined by $A_\mu=G_\mu\equiv\partial_\mu\ln\varphi^2$ since the projective symmetry is not broken there.

Finally, the on-shell action \eqref{eq:SOSR2F2} still enjoys the local gauged conformal symmetry, $S[g, A; \varphi]=S[\tilde{g}, \tilde{A}; \tilde{\varphi}]$, under the local gauged conformal transformations \eqref{eq:GCT2}, one can either directly gauge-fix $\varphi$ to $M_\mathrm{Pl}$ or choose a specific conformal factor $\Omega^2=\varphi^2/M_\mathrm{Pl}^2$ to conformally transform \eqref{eq:SOSR2F2} into the Einstein-frame action by $S[g_{\mu\nu}=\Omega^{-2}\tilde{g}_{\mu\nu}, A_\mu=\tilde{A}_\mu+\partial_\mu\ln\Omega^2 ;\varphi]$ as
\begin{align}\label{eq:SOSR2F2CT}
S[\tilde{g}, \tilde{A}]=\int\mathrm{d}^4x\sqrt{-\tilde{g}}&\left(\frac{M_\mathrm{Pl}^2}{2}\bar{R}(\tilde{g})-3M_\mathrm{Pl}^2\tilde{A}_\mu\tilde{A}^\mu\right.\nonumber\\
&\left.-\frac{1}{4\beta^2}F_{\mu\nu}(\tilde{A})^2-\frac{M_\mathrm{Pl}^4}{8\alpha}\right),
\end{align}
which is the Palatini Einstein gravity with a positive cosmological constant plus a Proca gauge field action. Fixing the gauge of $\varphi$ breaks the local gauge conformal symmetry of \eqref{eq:SOSR2F2}, and the would-be Goldstone field $G_\mu$ is therefore absorbed by $A_\mu$ to render a massive gauge field with a mass $m_A^2=6\beta^2M_\mathrm{Pl}^2$. When $A_\mu$ is decoupled below $m_A$, the metricity is deduced and the metric Einstein gravity with a positive cosmological constant is therefore recovered at this decoupling limit.

One can also arrive at the same result as \eqref{eq:SOSR2F2CT} from \eqref{eq:SR2F2varphi} by putting the connection on-shell after making either local conformal transformations  \eqref{eq:CT} or gauge-fixing $\varphi$ to $M_\mathrm{Pl}$. In specific, since the action \eqref{eq:SR2F2varphi}  is locally gauged conformal invariant,  we can fix the gauge of $\varphi$ to some constant scale $M_\mathrm{Pl}$, 
\begin{align}
S[g,\Gamma]=\int\mathrm{d}^4x\sqrt{-g}\left(\frac{M_\mathrm{Pl}^2}{2}R(g,\Gamma)-\frac{F_{\mu\nu}^2(A)}{4\beta^2}-\frac{M_\mathrm{Pl}^4}{8\alpha}\right),
\end{align}
which, after putting $\Gamma$ on-shell, reduces to the same form as  \eqref{eq:SOSR2F2CT} (but without over-tilde symbols). We can also choose a specific conformal factor $\Omega^2=\varphi^2/M_\mathrm{Pl}^2$ to conformally transform \eqref{eq:SR2F2varphi} into the Einstein-frame action by $S[g_{\mu\nu}=\Omega^{-2}\tilde{g}_{\mu\nu}, \Gamma^\rho_{\mu\nu}=\tilde{\Gamma}^\rho_{\mu\nu}; \varphi]$  as
\begin{align}
S[\tilde{g},\tilde{\Gamma}]=\int\mathrm{d}^4x\sqrt{-\tilde{g}}\left(\frac{M_\mathrm{Pl}^2}{2}R(\tilde{g},\tilde{\Gamma})-\frac{F_{\mu\nu}^2(\tilde{A})}{4\beta^2}-\frac{M_\mathrm{Pl}^4}{8\alpha}\right),
\end{align}
which, after putting $\tilde{\Gamma}$ on-shell, reduces to the same result as  \eqref{eq:SOSR2F2CT}. As also shown in Sec. \ref{subsec:R2}, this suggests that, starting from a locally gauged conformal invariant action, either gauge-fixing or a special local conformal transformation commutes with putting the connection on-shell. In what follows, we will only show one of these routines to simplify the discussions.

\section{Inclusion of matter field}\label{sec:addmatter}

Now that the Palatini quadratic gravity simply reproduces the metric Einstein gravity with a positive cosmological constant in a gauge-fixing form, we need to add matter field to the Palatini quadratic gravity in order to account for the inflaton field responsible for the cosmic inflation. There are two ways to add the matter field: either preserving or breaking the local gauged conformal symmetry.

\subsection{Preserving the local conformal symmetry}\label{subsec:preserve}

\subsubsection{Palatini $R^2$ gravity}

We start with the Palatini $R^2$ gravity with inclusion of a matter field $h$ as
\begin{align}\label{eq:SR2h}
S[g,\Gamma; h]=\int\mathrm{d}^4x\sqrt{-g}&\left(\frac{\alpha}{2}R^2+\frac{\xi h^2}{2}R-\frac12(D_\mu h)^2-V\right),
\end{align}
where the matter potential $V(h)\equiv(\lambda/4)h^4$ is defined without a dimensional scale, and the gauge covariant derivative is defined as $D_\mu=\nabla_\mu-\frac12A_\mu$  so that the gauge covariant derivative term transforms as $D_\mu h=\Omega(\tilde{D}_\mu\tilde{h})$ under $\tilde{h}=\Omega^{-1}h$ and  \eqref{eq:CT} that results in $D_\mu=\nabla_\mu-\frac12\tilde{A}_\mu-\frac12\partial_\mu\ln\Omega^2\equiv\tilde{D}_\mu-\partial_\mu\ln\Omega$. Therefore, the action \eqref{eq:SR2h} preserves the local gauged conformal symmetry, $S[g,\Gamma,h]=S[\tilde{g}, \tilde{\Gamma}, \tilde{h}]$, under the local gauged conformal transformations,
\begin{align}\label{eq:GCT3}
\tilde{g}_{\mu\nu}=\Omega(x)^2g_{\mu\nu}, \quad \tilde{\Gamma}^\rho_{\mu\nu}=\Gamma^\rho_{\mu\nu}, \quad \tilde{h}=\Omega^{-1}h,
\end{align}
where $\tilde{A}_\mu=A_\mu-\partial_\mu\ln\Omega^2$ is implicitly implied since it is not an independent transformation but as a result of the local conformal transformations \eqref{eq:CT}. Introducing the auxiliary field $\varphi$ to rewrite $\alpha^2R^2=\varphi^2R-\varphi^4/(4\alpha)$, one obtains the Jordan-frame action,
\begin{align}\label{eq:SR2hrho}
S[g,\Gamma; h, \rho]=\int\mathrm{d}^4x\sqrt{-g}\left(\frac{\rho^2}{2}R-\frac12(D_\mu h)^2-U(h,\rho)\right)
\end{align} 
with $\rho^2\equiv\varphi^2+\xi h^2$ and $U(h,\rho)\equiv(\lambda/4)h^4+(\rho^2-\xi h^2)^2/(8\alpha)$. 

Note that at this point $A_\mu$ is not an independent degree of freedom from the connection $\Gamma^\rho_{\mu\nu}$, thus one cannot separately vary the actions \eqref{eq:SR2h} or \eqref{eq:SR2hrho} with respect to $A_\mu$ from $\Gamma^\rho_{\mu\nu}$. In fact, $A_\mu$ only becomes an independent residual degree of freedom after putting the connection on-shell  due to the explicit presence of $A_\mu$ in the $D_\mu$ term that breaks the projective symmetry, 
\begin{align}
 S[g,\tilde{\Gamma},h,\rho]\neq S[g,\Gamma,h,\rho],\quad \tilde{\Gamma}^\rho_{\mu\nu}=\Gamma^\rho_{\mu\nu}+\delta^\rho_\mu\xi_\nu,
\end{align}
which would otherwise fix the Weyl gauge field $A_\mu$ from gauge-fixing the arbitrary vector field $\xi_\mu(x)$.
To put the connection on-shell, one first varies the action \eqref{eq:SR2hrho} with respect to the $\Gamma^\rho_{\mu\nu}$, and then obtain the EoM for the connection as
\begin{align}\label{eq:EOMGammaR2h}
\nabla_\lambda(\sqrt{-g}\rho^2g^{\mu\nu})-\nabla_\sigma(\sqrt{-g}\rho^2g^{\sigma(\mu})\delta^{\nu)}_\lambda=\frac{h}{2}\sqrt{-g}\delta^{(\mu}_\lambda D^{\nu)} h,
\end{align}
which, after contracting $\lambda=\nu$ and then plugging back the gauge covariant derivative term, becomes $\nabla_\lambda(\sqrt{-g}\rho^2g^{\mu\nu})-\frac25\nabla_\sigma(\sqrt{-g}\rho^2g^{\sigma(\mu})\delta^{\nu)}_\lambda=0$. This returns back the same solution as \eqref{eq:GammaR2F2}, and then the action \eqref{eq:SR2hrho} with on-shell connection reads
\begin{align}\label{eq:SR2hrhoOS}
S[g, A; h,\rho]=&\int\mathrm{d}^4x\sqrt{-g}\left(\frac{\rho^2}{2}\bar{R}(g)+3(\bar{\nabla}_\mu\rho)^2-U(h,\rho)\right.\nonumber\\
&\left.-3\rho^2(\partial_\mu\ln\rho^2-A_\mu)^2-\frac12(D_\mu h)^2\right),
\end{align}
which also enjoys a local gauged conformal symmetry, $S[g; h, \rho]=S[\tilde{g}; \tilde{h}, \tilde{\rho}]$, under the local gauged conformal transformations,
\begin{align}\label{eq:GCT5}
\tilde{g}_{\mu\nu}=\Omega^2g_{\mu\nu}, \quad \tilde{h}=\Omega^{-1}h, \quad \tilde{\rho}=\Omega^{-1}\rho.
\end{align}

This gauge symmetry allows us to fix one of the scalar degrees of freedom, for example, gauge-fixing $\rho$ to the Planck scale $M_\mathrm{Pl}$, and then the action \eqref{eq:SR2hrhoOS} reduces to
\begin{align}\label{eq:SR2hrhoOSGFrho}
S[g, A, h]=\int\mathrm{d}^4x&\sqrt{-g}\left(\frac{M_\mathrm{Pl}^2}{2}\bar{R}(g)-3M_\mathrm{Pl}^2A_\mu^2\right.\nonumber\\
&\left.-\frac12(D_\mu h)^2-U(h,M_\mathrm{Pl})\right),
\end{align}
Due to the absence of kinetic term for $A_\mu$,  it can be integrated out by putting it on-shell via its EoM (a constraint equation),
\begin{align}
A_\mu=\frac{\partial_\mu\ln h^2}{1+24M_\mathrm{Pl}^2/h^2},
\end{align}
and then the action further reduces to
\begin{align}
S[g,h]=\int\mathrm{d}^4x\sqrt{-g}&\left(\frac{M_\mathrm{Pl}^2}{2}\bar{R}(g)-\frac12\frac{(\partial_\mu h)^2}{1+\frac{h^2}{24M_\mathrm{Pl}^2}}\right.\nonumber\\
&\left.\qquad\qquad\qquad-U(h,M_\mathrm{Pl})\right),
\end{align}
which, after normalizing the kinetic term by redefining
\begin{align}
h=2\sqrt{6}M_\mathrm{Pl}\sinh\frac{\phi}{2\sqrt{6}M_\mathrm{Pl}},
\end{align}
becomes
\begin{align}\label{eq:SR2hrhoOSGFrhoOSA}
S[g,\phi]=\int\mathrm{d}^4x\sqrt{-g}\left(\frac{M_\mathrm{Pl}^2}{2}\bar{R}(g)-\frac12(\partial_\mu\phi)^2-W(\phi)\right)
\end{align}
with the potential $U(h(\phi),M_\mathrm{Pl})$ abbreviated as $W(\phi)$ of form
\begin{align}
W(\phi)=\frac{M_\mathrm{Pl}^4}{8\alpha}&\left(1-24\xi\sinh^2\frac{\phi}{2\sqrt{6}M_\mathrm{Pl}}\right)^2\nonumber\\
&+144\lambda M_\mathrm{Pl}^4\sinh^4\frac{\phi}{2\sqrt{6}M_\mathrm{Pl}}.
\end{align}

If all effective potential terms of the matter field are absent (namely $\xi=0$ and $\lambda=0$), the final reduced theory is the metric Einstein gravity with a cosmological constant. Otherwise, the potential $W(\phi)$ does not admit an asymptotically flat potential since $W(\phi)$ is divergent at $\phi\rightarrow\infty$ limit. In fact, $W(\phi)$ supports a small-field  tachyonic  inflation at small $\phi$ limit for hierarchical couplings $\alpha\lambda\ll\xi^2\ll1$ with the approximated potential
\begin{align}
\frac{W(\phi)}{M_\mathrm{Pl}^4/8\alpha}\approx\left(1-24\xi\sinh^2\frac{\phi}{2\sqrt{6}M_\mathrm{Pl}}\right)^2\approx1-\frac{2\xi}{M_\mathrm{Pl}^2}\phi^2,
\end{align}
from which the slow-roll parameters can be expanded as
\begin{align}
\epsilon&=\frac{M_\mathrm{Pl}^2}{2}\left(\frac{W'(\phi)}{W(\phi)}\right)^2=48\xi^2\sinh^2\frac{\phi}{\sqrt{6}M_\mathrm{Pl}}+\mathcal{O}(\xi^3),\\
\eta&=M_\mathrm{Pl}^2\frac{W''(\phi)}{W(\phi)}=-4\xi\cosh\frac{\phi}{\sqrt{6}M_\mathrm{Pl}}+\mathcal{O}(\xi^2).
\end{align}
Therefore, the consistency relation $r=-8n_t$ is unchanged but the scalar spectral index $n_s=1+2\eta_*-6\epsilon_*$ and the tensor-to-scalar ratio $r=16\epsilon_*$ evaluated at the horizon crossing of some pivot scale $k_*=a(t_*)H(t_*)$ is related by
\begin{align}
r=12(1-n_s)^2+\mathcal{O}(\xi^2).
\end{align}

\subsubsection{Palatini $R^2+R_{[\mu\nu]}^2$ gravity}

Parallel discussions also apply for Palatini $R^2+R_{[\mu\nu]}^2$ gravity with an action of form
\begin{align}\label{eq:SR2F2h}
S[g, \Gamma; h]=\int\mathrm{d}^4&x\sqrt{-g}\left(\frac{\alpha}{2}R(g,\Gamma)^2-\frac{1}{4\beta^2}R_{[\mu\nu]}(\Gamma)^2\right.\nonumber\\
&\left.+\frac{\xi h^2}{2}R-\frac12(D_\mu h)^2-V(h)\right),
\end{align}
which, after replacing $\alpha^2 R^2=\varphi^2R-\varphi^4/(4\alpha)$, becomes
\begin{align}\label{eq:SR2F2hrho}
S[g, \Gamma; h, \rho]=\int\mathrm{d}^4x&\sqrt{-g}\left(\frac{\rho^2}{2}R-\frac{F_{\mu\nu}^2}{4\beta^2}-\frac{(D_\mu h)^2}{2}-U\right)
\end{align}
with $\rho^2\equiv\varphi^2+\xi h^2$ and $U(h,\rho)\equiv(\lambda/4)h^4+(\rho^2-\xi h^2)/(8\alpha)$ as defined before. To put connection on-shell, solving the EoM of the connection from the action \eqref{eq:SR2F2hrho},
\begin{align}
\nabla_\lambda&(\sqrt{-g}\rho^2 g^{\mu\nu})-\nabla_\sigma(\sqrt{-g}\rho^2g^{\rho(\mu})\delta^{\nu)}_\lambda\nonumber\\
&=\frac{1}{\beta^2}\nabla_\sigma(\sqrt{-g}F^{\sigma(\mu})\delta^{\nu)}_\lambda-\frac{h}{2}\sqrt{-g}D^{(\mu}h\delta^{\nu)}_\lambda,
\end{align}
admits the same solution as \eqref{eq:GammaR2F2},  and the action  \eqref{eq:SR2F2hrho} with on-shell connection becomes
\begin{align}\label{eq:SR2F2hrhoOS}
&S[g, A; h, \rho]=\int\mathrm{d}^4x\sqrt{-g}\left(\frac{\rho^2}{2}\bar{R}+3(\bar{\nabla}_\mu\rho)^2-\frac{F_{\mu\nu}^2}{4\beta^2}\right.\nonumber\\
&\left.-3\rho^2(\partial_\mu\ln\rho^2-A_\mu)^2-\frac12(D_\mu h)^2-U(h,\rho)\right),
\end{align}
which still enjoys a local gauged conformal symmetry, $S[g, A; h, \rho]=S[\tilde{g}, \tilde{A}; \tilde{h}, \tilde{\rho}]$, under the local gauged conformal transformation \eqref{eq:GCT5}.  
Again, this allows us to gauge-fix $\rho$ to $M_\mathrm{Pl}$, and the reduced action reads
\begin{align}
S[g,A,h]=\int\mathrm{d}^4x&\sqrt{-g}\left(\frac{M_\mathrm{Pl}^2}{2}\bar{R}(g)-\frac{F_{\mu\nu}^2}{4\beta^2}-3M_\mathrm{Pl}^2A_\mu^2\right.\nonumber\\
&\left.-\frac12(D_\mu h)^2-U(h,M_\mathrm{Pl})\right).
\end{align}
This is the same action proposed in \cite{Ghilencea:2020piz} for the Palatini $R^2$ inflation with the same small-field tachyonic inflationary feature as \eqref{eq:SR2hrhoOSGFrhoOSA}. In a short summary, when the matter field is added in a way to preserve the local conformal symmetry (usually also break the projective symmetry at the same time), the asymptotically flat inflationary potential is not implied.

\subsection{Breaking the local conformal symmetry}

\subsubsection{Palatini $R^2$ gravity}

To break the local gauged conformal symmetry, we propose to replace the gauge covariant derivative $D_\mu$ in \eqref{eq:SR2F2h} with a normal covariant derivative $\nabla_\mu$, namely.
\begin{align}\label{eq:BSR2h}
S[g,\Gamma, h]=\int\mathrm{d}^4x\sqrt{-g}&\left(\frac{\alpha^2}{2}R(g,\Gamma)^2+\frac{G(h)}{2}R(g, \Gamma)\right.\nonumber\\
&\left.-\frac12(\nabla_\mu h)^2-V(h)\right).
\end{align}
As we will see shortly below that the cosmic inflation with an asymptotically flat potential is always obtained if one further breaks the local gauged conformal symmetry in the non-minimal coupling or matter potential  by adding lower-than-quadratic terms beyond $G(h)=\xi h^2$  or higher-than-quartic terms beyond $V(h)=(\lambda/4)h^4$ so that the ratio $V(h)/G(h)^2$ is an increasing function of $h$ at a large $h$ limit.

Similar to the previous sections, we first extract the scalar degree of freedom in the $R^2$ term by replacing $\alpha^2R^2=\varphi^2R-\varphi^4/(4\alpha)$,  then we obtain the Jordan-frame action
\begin{align}\label{eq:BSR2hrho}
S[g, \Gamma, h, \rho]=\int\mathrm{d}^4x\sqrt{-g}\left(\frac{\rho^2}{2}R-\frac{(\nabla_\mu h)^2}{2}-U\right)
\end{align}
with $\rho^2\equiv\varphi^2+G(h)$ and $U(h,\rho)\equiv V(h)+(\rho^2-G(h))^2/(8\alpha)$. Since the matter part of above action contains no connection-dependence, putting the connection on-shell simply returns back the solution \eqref{eq:GammaR2} with $A_\mu=G_\mu^{(\rho)}\equiv\partial_\mu\ln\rho^2$, which is consistent with the presence of projective symmetry of \eqref{eq:BSR2h} and \eqref{eq:BSR2hrho}.  Then, the action \eqref{eq:BSR2hrho} with on-shell connection reduces into
\begin{align}\label{eq:BSR2hrhoOS}
S[g, h, \rho]=\int\mathrm{d}^4x\sqrt{-g}\left(\frac{\rho^2}{2}\bar{R}+3(\bar{\nabla}_\mu\rho)^2-\frac{(\bar{\nabla}_\mu h)^2}{2}-U\right).
\end{align}
We next choose a special conformal factor $\Omega^2=\rho^2/M_\mathrm{Pl}^2$ as before to conformally transform the action \eqref{eq:BSR2hrhoOS} via $S[g_\mathrm{\mu\nu}=\Omega^{-2}\tilde{g}_{\mu\nu}; h, \rho]$ into
\begin{align}\label{eq:BSR2hrhoOSCT}
S[\tilde{g}, h, \rho]=\int\mathrm{d}^4x\sqrt{-\tilde{g}}\left(\frac{M_\mathrm{Pl}^2}{2}\bar{R}(\tilde{g})-\frac{(\tilde{\partial}_\mu h)^2}{2\Omega^2}-\frac{U}{\Omega^4}\right),
\end{align}
where $\rho$ admits no kinetic term but a constraint equation
\begin{align}
\rho^2=\frac{M_\mathrm{Pl}^2(G^2+8\alpha V)}{M_\mathrm{Pl}^2G+4\alpha X}
\end{align}
with the abbreviation $X\equiv-\frac12(\partial_\mu h)^2$ for the matter kinetic term. Hereafter we will rename $\tilde{g}_{\mu\nu}$ as $g_{\mu\nu}$ and drop the tilde symbol henceforth for simplicity.

Since $\rho$ is not a dynamical degree of freedom, we can integrate it out by putting the $\rho$ field on-shell with above constraint equation,  and the final result is simply a K-essence theory \cite{Armendariz-Picon:1999hyi,Armendariz-Picon:2000nqq,Armendariz-Picon:2000ulo,Gialamas:2019nly},
\begin{align}\label{eq:Kessence}
S[g,h]=\int\mathrm{d}^4\sqrt{-\tilde{g}}\left(\frac{M_\mathrm{Pl}^2}{2}\bar{R}+KX+L\frac{X^2}{M_\mathrm{Pl}^4}-W\right),
\end{align}
where some dimensionless abbreviations are defined as
\begin{align}
\hat{G}\equiv\frac{G(h)}{M_\mathrm{Pl}^2}, &\quad \hat{V}\equiv\frac{V(h)}{M_\mathrm{Pl}^4}, \\
K\equiv\frac{\hat{G}}{\hat{G}^2+8\alpha\hat{V}}=\frac{\hat{G}L}{2\alpha}, &\quad W\equiv\frac{M_\mathrm{Pl}^4\hat{V}}{\hat{G}^2+8\alpha\hat{V}}.
\end{align}
If both the non-minimal coupling $G=\xi h^2$ and the matter potential $V=(\lambda/4)h^4$ include no extra dimensional scales, then the effective potential $W$ is merely a cosmological constant,
\begin{align}
W=\frac{\lambda M_\mathrm{Pl}^4}{4(\xi^2+2\alpha\lambda)}.
\end{align}
However, if $G(h)$ or $V(h)$ is amended with additional dimensional scales to break the local gauged conformal symmetry in such a way that $G$ contains lower-than-quadratic terms, or $V$ contains higher-than-quartic terms,
\begin{align}
G(h)&=\xi h^2+\sum\limits_{n<2}\xi_n\Lambda_n^2\left(\frac{h}{\Lambda_n}\right)^n,\\
V(h)&=\frac{\lambda}{4}h^4+\sum\limits_{n>4}\lambda_n\Lambda_n^4\left(\frac{h}{\Lambda_n}\right)^n
\end{align}
so that $\hat{G}^2/\hat{V}$ is a decreasing function of $h$ at large $h$ (as first observed in \cite{Enckell:2018hmo} for a particular example with $G(h)=M_\mathrm{Pl}^2+\xi h^2$), then the effective potential $W$ always admits an asymptotically flat behavior,
\begin{align}
W=\frac{M_\mathrm{Pl}^4/8\alpha}{\left(1+\frac{\hat{G}^2}{8\alpha\hat{V}}\right)}\approx\frac{M_\mathrm{Pl}^4}{8\alpha}\left(1-\frac{\hat{G}^2}{8\alpha\hat{V}}\right).
\end{align}
Note that the inflationary potential $W$ is even more flattened when the potential $V$ becomes very steep. Therefore, this k-inflation \cite{Armendariz-Picon:1999hyi,Garriga:1999vw} but with an asymptotically flat potential largely emerges as a result of the broken local gauged conformal symmetry in both matter kinetic and potential terms (regarding the non-minimal coupling term as some kind of effective potential term induced by the background gravity).

The inflationary cosmology for the above K-essence action \eqref{eq:Kessence} has been already worked out for a general $P(X)$ theory in the second-order perturbation theory \cite{Martin:2013uma}. For our particular model with $P(X,h)=KX+LX^2/M_\mathrm{Pl}^4-W(h)$, the Einstein equation and scalar EoM,
\begin{align}
M_\mathrm{Pl}^2G_{\mu\nu}&=T_{\mu\nu}\equiv Pg_{\mu\nu}+P_X\nabla_\mu h\nabla_\nu h,\\
\sqrt{-g}P_h&+\nabla_\mu(\sqrt{-g}P_X\nabla^\mu h)=0,
\end{align}
can be expressed in the Friedmann-Lema\^{i}tre-Robertson-Walker (FLRW) metric for a background spatial homogeneous scalar field $h(t)$ as
\begin{align}
3M_\mathrm{Pl}^2H^2&=\frac12\left(K+\frac32L\frac{\dot{h}^2}{M_\mathrm{Pl}^4}\right)\dot{h}^2+W,\\
M_\mathrm{Pl}^2\dot{H}&=-\frac12\left(K+L\frac{\dot{h}^2}{M_\mathrm{Pl}^4}\right)\dot{h}^2,
\end{align}
and
\begin{align}
&\left(K+3L\frac{\dot{h}^2}{M_\mathrm{Pl}^4}\right)\ddot{h}+3H\left(K+L\frac{\dot{h}^2}{M_\mathrm{Pl}^4}\right)\dot{h}\nonumber\\
&\qquad+\frac12\left(K'+\frac32L'\frac{\dot{h}^2}{M_\mathrm{Pl}^4}\right)\dot{h}^2+W'(h)=0
\end{align}
respectively, and the sound speed $c_s^2=P_X/\rho_X$ with $\rho=2XP_X-P$ reads
\begin{align}
c_s^2=\frac{P_X}{P_X+2XP_{XX}}=\frac{K+L\dot{h}^2/M_\mathrm{Pl}^4}{K+3L\dot{h}^2/M_\mathrm{Pl}^4}.
\end{align}
The scalar and tensor power spectra to the first order are characterized by a series of Hubble flow and sound flow parameters,
\begin{align}
\epsilon_1=-\frac{\dot{H}}{H^2}, &\quad  \epsilon_{n+1}=\frac{\mathrm{d}\ln\epsilon_n}{\mathrm{d}N},\\
\delta_1=\frac{\dot{c}_s}{c_sH}, &\quad  \delta_{n+1}=\frac{\mathrm{d}\ln\delta_n}{\mathrm{d}N},
\end{align}
as
\begin{align}
\mathcal{P}_\zeta(k)=\frac{H_*^2(18e^{-3})}{8\pi^2M_\mathrm{Pl}^2\epsilon_1^*c_s^*}&\left[1-2(1+C)\epsilon_1^*-C\epsilon_2^*+(2+C)\delta_1^*\right.\nonumber\\
&\left.-(2\epsilon_1^*+\epsilon_2^*+\delta_1^*)\ln\frac{k}{k_*}\right],\\
\mathcal{P}_t(k)=\frac{2H_*^2(18e^{-3})}{\pi^2M_\mathrm{Pl}^2}&\left[1-2(1+C-\ln c_s^*)\epsilon_1^*-2\epsilon_1^*\ln\frac{k}{k_*}\right],
\end{align}
respectively,  from which the scalar/tensor spectral indexes and tensor-to-scalar ratio evaluated at the horizon crossing moment of some pivot scale $k_*=a(t_*)H(t_*)/c_s(t_*)$ are obtained as
\begin{align}
n_s&=1-2\epsilon_1^*-\epsilon_2^*-\delta_1^*,\\
n_t&=-2\epsilon_1^*,\\
r&=16\epsilon_1^*c_s^*\left[1+2\epsilon_1^*\ln c_s^*+C\epsilon_2^*-(2+C)\delta_1^*\right],
\end{align}
respectively, with the constant abbreviation $C\equiv1/3-\ln3$. Assuming that our plateau-like potential renders the horizon crossing moment for the CMB pivot scale deep into the slow-roll regime where $L\dot{h}^2/KM_\mathrm{Pl}^4\ll1$, then $c_s\approx1$ and $\delta_1^*\approx0$, therefore, the consistency relation deviates from the standard canonical single-field slow-roll inflation as
\begin{align}
r\approx-8n_t[1+C(1-n_s+n_t)],
\end{align}
which can be used for testing our model in the future CMB observations. 

Our model contains infinitely many possibilities to break the local conformal symmetry in the effective potential terms . The simplest choice would be $G(h)=M_\mathrm{Pl}^2+\xi h^2$ and $V(h)=(\lambda/4)h^4$ as first discussed in Ref.~ \cite{Enckell:2018hmo}, where the slow-roll predictions for the scalar perturbation amplitude and the scalar spectral index  are found to be coincided with the case without $R^2$ term but the tensor-to-scalar ratio is suppressed with respect to the case without $R^2$ term. Here we want to compare it with the case without breaking the local conformal symmetry in the effective potential terms, namely, $G(h)=\xi h^2$ and $V(h)=(\lambda/4)h^4$, which leads to a constant effective potential $W$ and a static solution $\dot{h}=0$. To compare with such a case with a small deviation, we assume a slow-roll condition $L\dot{h}^2/KM_\mathrm{Pl}^4\ll1$ so that one can define a canonical field $\phi=\int\mathrm{d}h\, K(h)$ with
\begin{align}
K(\hat{h})=\frac{1+\xi\hat{h}^2}{(1+\xi\hat{h}^2)^2+2\alpha\lambda\hat{h}^4},\quad \hat{h}\equiv h/M_\mathrm{Pl}.
\end{align}
Therefore, the usual slow-roll parameters can be similarly defined from the potential
\begin{align}
\hat{W}(\hat{h})=\frac{\lambda}{4}\frac{\hat{h}^4}{(1+\xi\hat{h}^2)^2+2\alpha\lambda\hat{h}^4}
\end{align}
with respect to the $\phi$ field as
\begin{align}
\epsilon_V(\hat{h})&=\frac{8(1+\xi\hat{h}^2)}{(\xi^2+2\alpha\lambda)\hat{h}^6+2\xi\hat{h}^4+\hat{h}^2},\\
\eta_V(\hat{h})&=\frac{4}{\hat{h}^2}\left(-\frac{3+2\xi\hat{h}^2}{1+\xi\hat{h}^2}+\frac{6(1+\xi\hat{h}^2)}{(1+\xi\hat{h}^2)^2+2\alpha\lambda\hat{h}^4}\right),
\end{align}
and the e-folding number from some pivot scale $\hat{h}_*$ to the end of inflation defined by $\epsilon_V(\hat{h}_\mathrm{end})=1$ can be obtained with a rather simple form as
\begin{align}
N_*=\int_{\hat{h}\mathrm{end}}^{\hat{h}_*}\mathrm{d}\hat{h}\,K(\hat{h})\frac{\hat{W}(\hat{h})}{\hat{W}'(\hat{h})}=\frac18(\hat{h}_*^2-\hat{h}_\mathrm{end}^2).
\end{align}
The cosmological predictions for the  scalar spectral index, the tensor-to-scalar ratio, and scalar perturbation amplitude can be obtained as
\begin{align}
&n_s(\hat{h}_*)=1-6\epsilon_V(\hat{h}_*)+2\eta_V(\hat{h}_*)=1-\frac{16}{\hat{h}_*^2}-\frac{8}{\hat{h}_*^2(1+\xi\hat{h}_*^2)},\\
&r(\hat{h}_*)=16\epsilon_V(\hat{h}_*)=\frac{128(1+\xi\hat{h}_*^2)}{(\xi^2+2\alpha\lambda)\hat{h}_*^6+2\xi\hat{h}_*^4+\hat{h}_*^2}.
\end{align}
\begin{align}
A_s(\hat{h}_*)&=\frac{1}{24\pi^2}\frac{\hat{W}(\hat{h}_*)}{\epsilon_V(\hat{h}_*)}=\frac{\lambda\hat{h}_*^6}{768\pi^2(1+\xi\hat{h}_*^2)}.
\end{align}
After expanded in the leading order of $\hat{h}_\mathrm{end}$, the above cosmological predictions become
\begin{align}
n_s&=1-\frac{2}{N_*}-\frac{1}{N_*(1+8\xi N_*)}+\mathcal{O}(\hat{h}_\mathrm{end}^2),\\
r&=\frac{16(1+8\xi N_*)}{N_*+16\xi N_*^2+64N_*^3(\xi^2+2\alpha\lambda)}+\mathcal{O}(\hat{h}_\mathrm{end}^2),\\
A_s&=\frac{2\lambda N_*^3}{3\pi^2(1+8\xi N_*)}+\mathcal{O}(\hat{h}_\mathrm{end}^2).
\end{align}
It can be numerically checked that this approximation is sufficiently stable for model predictions, which are the functions of $N_*$ with input parameters $\lambda$, $\xi$, and $\alpha$. In order to identify the parameter regions of observational interest, we can use the measured values of $n_s$ and $A_s$ to fix  $\lambda$, $\xi$, 
\begin{align}
\xi&=\frac{(1-n_s)N_*-3}{16N_*-8N_*^2(1-n_s)},\\
\lambda&=\frac{3\pi^2A_s}{2N_*^4(1-n_s)-4N_*^3},
\end{align}
and then the tensor-to-scalar ratio reads
\begin{align}
r=\frac{32-16N_*(1-n_s)}{192\pi^2\alpha A_s[2-N_*(1-n_s)]-N_*}.
\end{align}
Requiring $r$ to be smaller than the current upper bound $r_{0.05}<0.036$~\cite{BICEP:2021xfz}, $\alpha$ should satisfy
\begin{align}
\alpha>\frac{32-16N_*(1-n_s-r_{0.05}/16)}{192\pi^2A_s[2-N_*(1-n_s)]r_{0.05}}.
\end{align}
Using the best-fit values $n_s=0.9649$ and $\ln(10^{10}A_s)=3.045$ from Planck 2018 TT,TE,EE+lowE constraints~\cite{Planck:2018jri}, we  finally identify the parameter space of $\alpha$ as
\begin{align}
\alpha>1.04\times10^8+\frac{4.08\times10^8}{56.98-N_*}.
\end{align}
The remaining freedom on $N_*$ can be traced back to different reheating histories. In general, as long as above condition on $\alpha$ is satisfied, one can always find the parameter regions for $\lambda$ and $\xi$ to simultaneously meet the observational constraints $n_s=0.9649$, $\ln(10^{10}A_s)=3.045$ and $r_{0.05}<0.036$.

\subsubsection{Palatini $R^2+R_{[\mu\nu]}^2$ gravity}

Parallel discussions also apply for Palatini $R^2+R_{[\mu\nu]}^2$ gravity with an action of form
\begin{align}\label{eq:BSR2F2h}
S[g, \Gamma, h]&=\int\mathrm{d}^4x\sqrt{-g}\left(\frac{\alpha}{2}R(g, \Gamma)^2-\frac{1}{4\beta^2}F_{\mu\nu}(A)^2\right.\nonumber\\
&\left.+\frac{G(h)}{2}R(g, \Gamma)-\frac12(\nabla_\mu h)^2-V(h)\right),
\end{align}
which, after replacing $\alpha^2R^2=\varphi^2R-\varphi^4/(4\alpha)$, becomes
\begin{align}\label{eq:BSR2F2hrho}
S[g, \Gamma, h, \rho]=\int\mathrm{d}^4x&\sqrt{-g}\left(\frac{\rho^2}{2}R(g,\Gamma)-\frac{1}{4\beta^2}F_{\mu\nu}(A)^2\right.\nonumber\\
&\left.-\frac12(\nabla_\mu h)^2-U(h, \rho)\right).
\end{align}
Putting the connection  on-shell with the same solution \eqref{eq:GammaR2F2} gives rise to an action of a form
\begin{align}
S[g, A, h, &\rho]=\int\mathrm{d}^4x\sqrt{-g}\left(\frac{\rho^2}{2}\bar{R}+3(\bar{\nabla}_\mu\rho)^2-\frac{1}{4\beta^2}F_{\mu\nu}^2\right.\nonumber\\
&\left.-3\rho^2(\partial_\mu\ln\rho^2-A_\mu)^2-\frac12(\bar{\nabla}_\mu h)^2-U\right),
\end{align}
which, after conformally transformed into Einstein frame via $S[g_{\mu\nu}=\Omega^{-2}\tilde{g}_{\mu\nu}, A_\mu=\tilde{A}_\mu+\partial_\mu\ln\Omega^2, h, \rho]$ with a specific conformal factor $\Omega^2=\rho^2/M_\mathrm{Pl}^2$, is reduced into
\begin{align}
S[\tilde{g}, \tilde{A}, h, \rho]=\int\mathrm{d}^4x\sqrt{-\tilde{g}}&\left(\frac{M_\mathrm{Pl}^2}{2}\tilde{\bar{R}}-\frac{\tilde{F}_{\mu\nu}^2}{4\beta^2}-3M_\mathrm{Pl}^2\tilde{A}_\mu^2\right.\nonumber\\
&\left.-\frac{1}{2\Omega^2}(\partial_\mu h)^2-\frac{U}{\Omega^4}\right).
\end{align}
When $A_\mu$ is decoupled below the scale $\sqrt{6}\beta M_\mathrm{Pl}$, we return back to \eqref{eq:BSR2hrhoOSCT} that immediately leads to the K-essence theory \eqref{eq:Kessence} and hence an asymptotically flat inflationary potential is similarly obtained. In a short summary, the asymptotically flat potential emerges as a result of breaking the local conformal symmetry appropriately for both scalar kinetic and effective potential terms, and is independent of the presence or absence of the projective symmetry as shown for \eqref{eq:BSR2h} or \eqref{eq:BSR2F2h}, respectively.

\section{Conclusions and discussions}\label{sec:conclusion}

Cosmic inflation is the standard pillar for the standard model of modern cosmology, describing a period of nearly exponential expansion of spacetime in the very early Universe to solve several fine-tuning problems of the standard hot big bang scenario and generate  nearly scale-invariant primordial perturbations observed in the cosmic microwave background and large scale structures. The current observational data prefers a single-field slow-roll plateau-like inflationary potential, which could be theoretically motivated from the cosmological attractor approach. A conformon field with a wrong-sign kinetic term is introduced to respect the local conformal symmetry and a second scalar field is added in such a way to impose an additional global symmetry jointed with the conformon field, which is broken by the potential term but with the local conformal symmetry intact.  After fixing the gauge of conformon field, the potential term with broken global symmetry gives rise to the exponentially flattened inflationary potential. 

However, this approach introduces the wrong-sign conformon field at the  price of introducing an additional global symmetry for inflationary model buildings. Nevertheless, the wrong-sign conformon field could naturally arise in the Palatini quadratic gravity, though an additional global symmetry is also adopted for inflationary model buildings. In this paper, we point out that, in Palatini quadratic gravity, such an encumbrance of an additional global symmetry is needless. Appropriately breaking the local conformal symmetry alone for both kinetic and potential terms of a matter field is sufficient to produce an asymptotically flat inflationary potential regardless of the high steepness of original matter potential. 

For future perspectives, it is still mysterious what position should we find for the Palatini quadratic gravity in approaching the underlying quantum gravity. A related question is that, for Palatini quadratic gravity without matter field or with conformally invariant matter field, since the local conformal symmetry is a gauge symmetry, then what causes this redundancy or what is the origin for this local conformal symmetry? This is a profound question \cite{Carlip:2012wa,Witten:2017hdv} on how gauge symmetry emerges from more physical symmetry \cite{Barcelo:2016xhp,Barcelo:2021idt}.

The next question concerns the transition from the local conformally symmetric matter phase to the locally conformal-symmetry broken matter phase. Breaking the local conformal symmetry in matter potential is easy by quantum corrections or renormalization group flow. However, the reduction of a gauge covariant derivative term into a normal covariant derivative term is unclear. A dynamical mechanism for triggering such a broken conformal symmetry in the kinetic term is desirable.

The last question runs into the initial conditions for the cosmic inflation, which is usually the realm of the quantum cosmology \cite{Halliwell:1989myn} for the no-boundary  \cite{Hartle:1983ai,Hawking:1983hj} and tunneling  \cite{Vilenkin:1982de,Linde:1983mx,Vilenkin:1984wp,Vilenkin:1986cy,Vilenkin:1987kf} proposals. As far as we know, there is currently no study on quantum cosmology starting from the Palatini quadratic gravity, which might be related to the recent new result \cite{Wang:2019spw} in presence of non-minimal coupling compared to the case of absence \cite{Vilenkin:2018dch,Vilenkin:2018oja}.

\begin{acknowledgments}
We thank Li Li, Run-Qiu Yang, Shan-Ming Ruan for helpful discussions.
This work is supported by the National Key Research and Development Program of China Grant No.2020YFC2201501, 
the National Natural Science Foundation of China Grants No.12105344, No. 11647601, No.11821505, No.11851302, No.12047503, No.11991052, No.12075297 and No. 12047558, 
the Key Research Program of the CAS Grant No. XDPB15, 
the Key Research Program of Frontier Sciences of CAS, 
and the Science Research Grants from the China Manned Space Project with NO. CMS-CSST-2021-B01.
\end{acknowledgments}


\bibliographystyle{utphys}
\bibliography{ref}

\providecommand{\href}[2]{#2}\begingroup\raggedright\begin{thebibliography}{10}

\bibitem{Brout:1977ix}
R.~Brout, F.~Englert, and E.~Gunzig, ``{The Creation of the Universe as a
  Quantum Phenomenon},''
  \href{http://dx.doi.org/10.1016/0003-4916(78)90176-8}{{\em Annals Phys.}
  {\bfseries 115} (1978) 78}.

\bibitem{Starobinsky:1980te}
A.~A. Starobinsky, ``{A New Type of Isotropic Cosmological Models Without
  Singularity},'' \href{http://dx.doi.org/10.1016/0370-2693(80)90670-X}{{\em
  Phys. Lett. B} {\bfseries 91} (1980) 99--102}.

\bibitem{Kazanas:1980tx}
D.~Kazanas, ``{Dynamics of the Universe and Spontaneous Symmetry Breaking},''
  \href{http://dx.doi.org/10.1086/183361}{{\em Astrophys. J. Lett.} {\bfseries
  241} (1980) L59--L63}.

\bibitem{Sato:1980yn}
K.~Sato, ``{First Order Phase Transition of a Vacuum and Expansion of the
  Universe},'' {\em Mon. Not. Roy. Astron. Soc.} {\bfseries 195} (1981)
  467--479.

\bibitem{Guth:1980zm}
A.~H. Guth, ``{The Inflationary Universe: A Possible Solution to the Horizon
  and Flatness Problems},''
  \href{http://dx.doi.org/10.1103/PhysRevD.23.347}{{\em Phys. Rev. D}
  {\bfseries 23} (1981) 347--356}.

\bibitem{Linde:1981mu}
A.~D. Linde, ``{A New Inflationary Universe Scenario: A Possible Solution of
  the Horizon, Flatness, Homogeneity, Isotropy and Primordial Monopole
  Problems},'' \href{http://dx.doi.org/10.1016/0370-2693(82)91219-9}{{\em Phys.
  Lett. B} {\bfseries 108} (1982) 389--393}.

\bibitem{Albrecht:1982wi}
A.~Albrecht and P.~J. Steinhardt, ``{Cosmology for Grand Unified Theories with
  Radiatively Induced Symmetry Breaking},''
  \href{http://dx.doi.org/10.1103/PhysRevLett.48.1220}{{\em Phys. Rev. Lett.}
  {\bfseries 48} (1982) 1220--1223}.

\bibitem{Linde:1983gd}
A.~D. Linde, ``{Chaotic Inflation},''
  \href{http://dx.doi.org/10.1016/0370-2693(83)90837-7}{{\em Phys. Lett. B}
  {\bfseries 129} (1983) 177--181}.

\bibitem{Mukhanov:1981xt}
V.~F. Mukhanov and G.~V. Chibisov, ``{Quantum Fluctuations and a Nonsingular
  Universe},'' {\em JETP Lett.} {\bfseries 33} (1981) 532--535.

\bibitem{Mukhanov:1982nu}
V.~F. Mukhanov and G.~V. Chibisov, ``{The Vacuum energy and large scale
  structure of the universe},'' {\em Sov. Phys. JETP} {\bfseries 56} (1982)
  258--265.

\bibitem{Hawking:1982cz}
S.~W. Hawking, ``{The Development of Irregularities in a Single Bubble
  Inflationary Universe},''
  \href{http://dx.doi.org/10.1016/0370-2693(82)90373-2}{{\em Phys. Lett. B}
  {\bfseries 115} (1982) 295}.

\bibitem{Guth:1982ec}
A.~H. Guth and S.~Y. Pi, ``{Fluctuations in the New Inflationary Universe},''
  \href{http://dx.doi.org/10.1103/PhysRevLett.49.1110}{{\em Phys. Rev. Lett.}
  {\bfseries 49} (1982) 1110--1113}.

\bibitem{Starobinsky:1982ee}
A.~A. Starobinsky, ``{Dynamics of Phase Transition in the New Inflationary
  Universe Scenario and Generation of Perturbations},''
  \href{http://dx.doi.org/10.1016/0370-2693(82)90541-X}{{\em Phys. Lett. B}
  {\bfseries 117} (1982) 175--178}.

\bibitem{Bardeen:1983qw}
J.~M. Bardeen, P.~J. Steinhardt, and M.~S. Turner, ``{Spontaneous Creation of
  Almost Scale - Free Density Perturbations in an Inflationary Universe},''
  \href{http://dx.doi.org/10.1103/PhysRevD.28.679}{{\em Phys. Rev. D}
  {\bfseries 28} (1983) 679}.

\bibitem{Kodama:1984ziu}
H.~Kodama and M.~Sasaki, ``{Cosmological Perturbation Theory},''
  \href{http://dx.doi.org/10.1143/PTPS.78.1}{{\em Prog. Theor. Phys. Suppl.}
  {\bfseries 78} (1984) 1--166}.

\bibitem{Mukhanov:1985rz}
V.~F. Mukhanov, ``{Gravitational Instability of the Universe Filled with a
  Scalar Field},'' {\em JETP Lett.} {\bfseries 41} (1985) 493--496.

\bibitem{WMAP:2012nax}
{\bfseries WMAP} Collaboration, G.~Hinshaw {\em et~al.}, ``{Nine-Year Wilkinson
  Microwave Anisotropy Probe (WMAP) Observations: Cosmological Parameter
  Results},'' \href{http://dx.doi.org/10.1088/0067-0049/208/2/19}{{\em
  Astrophys. J. Suppl.} {\bfseries 208} (2013) 19},
  \href{http://arxiv.org/abs/1212.5226}{{\ttfamily arXiv:1212.5226
  [astro-ph.CO]}}.

\bibitem{Planck:2018jri}
{\bfseries Planck} Collaboration, Y.~Akrami {\em et~al.}, ``{Planck 2018
  results. X. Constraints on inflation},''
  \href{http://dx.doi.org/10.1051/0004-6361/201833887}{{\em Astron. Astrophys.}
  {\bfseries 641} (2020) A10},
  \href{http://arxiv.org/abs/1807.06211}{{\ttfamily arXiv:1807.06211
  [astro-ph.CO]}}.

\bibitem{BOSS:2016wmc}
{\bfseries BOSS} Collaboration, S.~Alam {\em et~al.}, ``{The clustering of
  galaxies in the completed SDSS-III Baryon Oscillation Spectroscopic Survey:
  cosmological analysis of the DR12 galaxy sample},''
  \href{http://dx.doi.org/10.1093/mnras/stx721}{{\em Mon. Not. Roy. Astron.
  Soc.} {\bfseries 470} no.~3, (2017) 2617--2652},
  \href{http://arxiv.org/abs/1607.03155}{{\ttfamily arXiv:1607.03155
  [astro-ph.CO]}}.

\bibitem{eBOSS:2020yzd}
{\bfseries eBOSS} Collaboration, S.~Alam {\em et~al.}, ``{Completed SDSS-IV
  extended Baryon Oscillation Spectroscopic Survey: Cosmological implications
  from two decades of spectroscopic surveys at the Apache Point Observatory},''
  \href{http://dx.doi.org/10.1103/PhysRevD.103.083533}{{\em Phys. Rev. D}
  {\bfseries 103} no.~8, (2021) 083533},
  \href{http://arxiv.org/abs/2007.08991}{{\ttfamily arXiv:2007.08991
  [astro-ph.CO]}}.

\bibitem{BICEPKeck:2021gln}
{\bfseries BICEP/Keck} Collaboration, P.~A.~R. Ade {\em et~al.}, ``{Improved
  Constraints on Primordial Gravitational Waves using Planck, WMAP, and
  BICEP/Keck Observations through the 2018 Observing Season},''
  \href{http://dx.doi.org/10.1103/PhysRevLett.127.151301}{{\em Phys. Rev.
  Lett.} {\bfseries 127} no.~15, (2021) 151301},
  \href{http://arxiv.org/abs/2110.00483}{{\ttfamily arXiv:2110.00483
  [astro-ph.CO]}}.

\bibitem{Bezrukov:2007ep}
F.~L. Bezrukov and M.~Shaposhnikov, ``{The Standard Model Higgs boson as the
  inflaton},'' \href{http://dx.doi.org/10.1016/j.physletb.2007.11.072}{{\em
  Phys. Lett. B} {\bfseries 659} (2008) 703--706},
  \href{http://arxiv.org/abs/0710.3755}{{\ttfamily arXiv:0710.3755 [hep-th]}}.

\bibitem{Galante:2014ifa}
M.~Galante, R.~Kallosh, A.~Linde, and D.~Roest, ``{Unity of Cosmological
  Inflation Attractors},''
  \href{http://dx.doi.org/10.1103/PhysRevLett.114.141302}{{\em Phys. Rev.
  Lett.} {\bfseries 114} no.~14, (2015) 141302},
  \href{http://arxiv.org/abs/1412.3797}{{\ttfamily arXiv:1412.3797 [hep-th]}}.

\bibitem{Kallosh:2013hoa}
R.~Kallosh and A.~Linde, ``{Universality Class in Conformal Inflation},''
  \href{http://dx.doi.org/10.1088/1475-7516/2013/07/002}{{\em JCAP} {\bfseries
  07} (2013) 002}, \href{http://arxiv.org/abs/1306.5220}{{\ttfamily
  arXiv:1306.5220 [hep-th]}}.

\bibitem{Kallosh:2013daa}
R.~Kallosh and A.~Linde, ``{Multi-field Conformal Cosmological Attractors},''
  \href{http://dx.doi.org/10.1088/1475-7516/2013/12/006}{{\em JCAP} {\bfseries
  12} (2013) 006}, \href{http://arxiv.org/abs/1309.2015}{{\ttfamily
  arXiv:1309.2015 [hep-th]}}.

\bibitem{Ferrara:2013rsa}
S.~Ferrara, R.~Kallosh, A.~Linde, and M.~Porrati, ``{Minimal Supergravity
  Models of Inflation},''
  \href{http://dx.doi.org/10.1103/PhysRevD.88.085038}{{\em Phys. Rev. D}
  {\bfseries 88} no.~8, (2013) 085038},
  \href{http://arxiv.org/abs/1307.7696}{{\ttfamily arXiv:1307.7696 [hep-th]}}.

\bibitem{Kallosh:2013yoa}
R.~Kallosh, A.~Linde, and D.~Roest, ``{Superconformal Inflationary
  $\alpha$-Attractors},'' \href{http://dx.doi.org/10.1007/JHEP11(2013)198}{{\em
  JHEP} {\bfseries 11} (2013) 198},
  \href{http://arxiv.org/abs/1311.0472}{{\ttfamily arXiv:1311.0472 [hep-th]}}.

\bibitem{Linde:2015uga}
A.~Linde, ``{Single-field $\alpha$-attractors},''
  \href{http://dx.doi.org/10.1088/1475-7516/2015/05/003}{{\em JCAP} {\bfseries
  05} (2015) 003}, \href{http://arxiv.org/abs/1504.00663}{{\ttfamily
  arXiv:1504.00663 [hep-th]}}.

\bibitem{Cai:2014bda}
Y.-F. Cai, J.-O. Gong, and S.~Pi, ``{Inflation beyond T-models and primordial
  B-modes},'' \href{http://dx.doi.org/10.1016/j.physletb.2014.09.009}{{\em
  Phys. Lett. B} {\bfseries 738} (2014) 20--24},
  \href{http://arxiv.org/abs/1404.2560}{{\ttfamily arXiv:1404.2560 [hep-th]}}.

\bibitem{Kallosh:2013tua}
R.~Kallosh, A.~Linde, and D.~Roest, ``{Universal Attractor for Inflation at
  Strong Coupling},''
  \href{http://dx.doi.org/10.1103/PhysRevLett.112.011303}{{\em Phys. Rev.
  Lett.} {\bfseries 112} no.~1, (2014) 011303},
  \href{http://arxiv.org/abs/1310.3950}{{\ttfamily arXiv:1310.3950 [hep-th]}}.

\bibitem{Giudice:2014toa}
G.~F. Giudice and H.~M. Lee, ``{Starobinsky-like inflation from induced
  gravity},'' \href{http://dx.doi.org/10.1016/j.physletb.2014.04.020}{{\em
  Phys. Lett. B} {\bfseries 733} (2014) 58--62},
  \href{http://arxiv.org/abs/1402.2129}{{\ttfamily arXiv:1402.2129 [hep-ph]}}.

\bibitem{Pallis:2013yda}
C.~Pallis, ``{Linking Starobinsky-Type Inflation in no-Scale Supergravity to
  MSSM},'' \href{http://dx.doi.org/10.1088/1475-7516/2014/04/024}{{\em JCAP}
  {\bfseries 04} (2014) 024}, \href{http://arxiv.org/abs/1312.3623}{{\ttfamily
  arXiv:1312.3623 [hep-ph]}}. [Erratum: JCAP 07, E01 (2017)].

\bibitem{Pallis:2014dma}
C.~Pallis, ``{Induced-Gravity Inflation in no-Scale Supergravity and Beyond},''
  \href{http://dx.doi.org/10.1088/1475-7516/2014/08/057}{{\em JCAP} {\bfseries
  08} (2014) 057}, \href{http://arxiv.org/abs/1403.5486}{{\ttfamily
  arXiv:1403.5486 [hep-ph]}}.

\bibitem{Pallis:2014boa}
C.~Pallis, ``{Reconciling Induced-Gravity Inflation in Supergravity With The
  Planck 2013 \& BICEP2 Results},''
  \href{http://dx.doi.org/10.1088/1475-7516/2014/10/058}{{\em JCAP} {\bfseries
  10} (2014) 058}, \href{http://arxiv.org/abs/1407.8522}{{\ttfamily
  arXiv:1407.8522 [hep-ph]}}.

\bibitem{Kallosh:2014rha}
R.~Kallosh, ``{More on Universal Superconformal Attractors},''
  \href{http://dx.doi.org/10.1103/PhysRevD.89.087703}{{\em Phys. Rev. D}
  {\bfseries 89} no.~8, (2014) 087703},
  \href{http://arxiv.org/abs/1402.3286}{{\ttfamily arXiv:1402.3286 [hep-th]}}.

\bibitem{Kallosh:2000ve}
R.~Kallosh, L.~Kofman, A.~D. Linde, and A.~Van~Proeyen, ``{Superconformal
  symmetry, supergravity and cosmology},''
  \href{http://dx.doi.org/10.1088/0264-9381/17/20/308}{{\em Class. Quant.
  Grav.} {\bfseries 17} (2000) 4269--4338},
  \href{http://arxiv.org/abs/hep-th/0006179}{{\ttfamily arXiv:hep-th/0006179}}.
  [Erratum: Class.Quant.Grav. 21, 5017 (2004)].

\bibitem{Ferrara:2010yw}
S.~Ferrara, R.~Kallosh, A.~Linde, A.~Marrani, and A.~Van~Proeyen, ``{Jordan
  Frame Supergravity and Inflation in NMSSM},''
  \href{http://dx.doi.org/10.1103/PhysRevD.82.045003}{{\em Phys. Rev. D}
  {\bfseries 82} (2010) 045003},
  \href{http://arxiv.org/abs/1004.0712}{{\ttfamily arXiv:1004.0712 [hep-th]}}.

\bibitem{Ferrara:2010in}
S.~Ferrara, R.~Kallosh, A.~Linde, A.~Marrani, and A.~Van~Proeyen,
  ``{Superconformal Symmetry, NMSSM, and Inflation},''
  \href{http://dx.doi.org/10.1103/PhysRevD.83.025008}{{\em Phys. Rev. D}
  {\bfseries 83} (2011) 025008},
  \href{http://arxiv.org/abs/1008.2942}{{\ttfamily arXiv:1008.2942 [hep-th]}}.

\bibitem{Kallosh:2013pby}
R.~Kallosh and A.~Linde, ``{Superconformal generalization of the chaotic
  inflation model $\frac{\lambda}{4} \phi^{4} - \frac{\xi}{2} \phi^{2}R$},''
  \href{http://dx.doi.org/10.1088/1475-7516/2013/06/027}{{\em JCAP} {\bfseries
  06} (2013) 027}, \href{http://arxiv.org/abs/1306.3211}{{\ttfamily
  arXiv:1306.3211 [hep-th]}}.

\bibitem{Kallosh:2013lkr}
R.~Kallosh and A.~Linde, ``{Superconformal generalizations of the Starobinsky
  model},'' \href{http://dx.doi.org/10.1088/1475-7516/2013/06/028}{{\em JCAP}
  {\bfseries 06} (2013) 028}, \href{http://arxiv.org/abs/1306.3214}{{\ttfamily
  arXiv:1306.3214 [hep-th]}}.

\bibitem{Hehl:1994ue}
F.~W. Hehl, J.~D. McCrea, E.~W. Mielke, and Y.~Ne'eman, ``{Metric affine gauge
  theory of gravity: Field equations, Noether identities, world spinors, and
  breaking of dilation invariance},''
  \href{http://dx.doi.org/10.1016/0370-1573(94)00111-F}{{\em Phys. Rept.}
  {\bfseries 258} (1995) 1--171},
  \href{http://arxiv.org/abs/gr-qc/9402012}{{\ttfamily arXiv:gr-qc/9402012}}.

\bibitem{Gronwald:1997bx}
F.~Gronwald, ``{Metric affine gauge theory of gravity. 1. Fundamental structure
  and field equations},''
  \href{http://dx.doi.org/10.1142/S0218271897000157}{{\em Int. J. Mod. Phys. D}
  {\bfseries 6} (1997) 263--304},
  \href{http://arxiv.org/abs/gr-qc/9702034}{{\ttfamily arXiv:gr-qc/9702034}}.

\bibitem{Ghilencea:2020piz}
D.~M. Ghilencea, ``{Palatini quadratic gravity: spontaneous breaking of gauged
  scale symmetry and inflation},''
  \href{http://dx.doi.org/10.1140/epjc/s10052-020-08722-0}{{\em Eur. Phys. J.
  C} {\bfseries 80} no.~12, (4, 2020) 1147},
  \href{http://arxiv.org/abs/2003.08516}{{\ttfamily arXiv:2003.08516
  [hep-th]}}.

\bibitem{Ghilencea:2018dqd}
D.~M. Ghilencea, ``{Spontaneous breaking of Weyl quadratic gravity to Einstein
  action and Higgs potential},''
  \href{http://dx.doi.org/10.1007/JHEP03(2019)049}{{\em JHEP} {\bfseries 03}
  (2019) 049}, \href{http://arxiv.org/abs/1812.08613}{{\ttfamily
  arXiv:1812.08613 [hep-th]}}.

\bibitem{Ghilencea:2019jux}
D.~M. Ghilencea, ``{Stueckelberg breaking of Weyl conformal geometry and
  applications to gravity},''
  \href{http://dx.doi.org/10.1103/PhysRevD.101.045010}{{\em Phys. Rev. D}
  {\bfseries 101} no.~4, (2020) 045010},
  \href{http://arxiv.org/abs/1904.06596}{{\ttfamily arXiv:1904.06596
  [hep-th]}}.

\bibitem{Ghilencea:2019rqj}
D.~M. Ghilencea, ``{Weyl R$^{2}$ inflation with an emergent Planck scale},''
  \href{http://dx.doi.org/10.1007/JHEP10(2019)209}{{\em JHEP} {\bfseries 10}
  (2019) 209}, \href{http://arxiv.org/abs/1906.11572}{{\ttfamily
  arXiv:1906.11572 [gr-qc]}}.

\bibitem{Ghilencea:2020rxc}
D.~M. Ghilencea, ``{Gauging scale symmetry and inflation: Weyl versus Palatini
  gravity},'' \href{http://dx.doi.org/10.1140/epjc/s10052-021-09226-1}{{\em
  Eur. Phys. J. C} {\bfseries 81} no.~6, (2021) 510},
  \href{http://arxiv.org/abs/2007.14733}{{\ttfamily arXiv:2007.14733
  [hep-th]}}.

\bibitem{Ghilencea:2021lpa}
D.~M. Ghilencea, ``{Standard Model in Weyl conformal geometry},''
  \href{http://arxiv.org/abs/2104.15118}{{\ttfamily arXiv:2104.15118
  [hep-ph]}}.

\bibitem{Ghilencea:2021jjl}
D.~M. Ghilencea and T.~Harko, ``{Cosmological evolution in Weyl conformal
  geometry},'' \href{http://arxiv.org/abs/2110.07056}{{\ttfamily
  arXiv:2110.07056 [gr-qc]}}.

\bibitem{Tang:2018mhn}
Y.~Tang and Y.-L. Wu, ``{Inflation in gauge theory of gravity with local
  scaling symmetry and quantum induced symmetry breaking},''
  \href{http://dx.doi.org/10.1016/j.physletb.2018.07.048}{{\em Phys. Lett. B}
  {\bfseries 784} (2018) 163--168},
  \href{http://arxiv.org/abs/1805.08507}{{\ttfamily arXiv:1805.08507 [gr-qc]}}.

\bibitem{Tang:2019uex}
Y.~Tang and Y.-L. Wu, ``{Weyl Symmetry Inspired Inflation and Dark Matter},''
  \href{http://dx.doi.org/10.1016/j.physletb.2020.135320}{{\em Phys. Lett. B}
  {\bfseries 803} (2020) 135320},
  \href{http://arxiv.org/abs/1904.04493}{{\ttfamily arXiv:1904.04493
  [hep-ph]}}.

\bibitem{Tang:2019olx}
Y.~Tang and Y.-L. Wu, ``{Conformal $\alpha$-attractor inflation with Weyl gauge
  field},'' \href{http://dx.doi.org/10.1088/1475-7516/2020/03/067}{{\em JCAP}
  {\bfseries 03} (2020) 067}, \href{http://arxiv.org/abs/1912.07610}{{\ttfamily
  arXiv:1912.07610 [hep-ph]}}.

\bibitem{Tang:2020ovf}
Y.~Tang and Y.-L. Wu, ``{Weyl scaling invariant $R^2$ gravity for inflation and
  dark matter},'' \href{http://dx.doi.org/10.1016/j.physletb.2020.135716}{{\em
  Phys. Lett. B} {\bfseries 809} (2020) 135716},
  \href{http://arxiv.org/abs/2006.02811}{{\ttfamily arXiv:2006.02811
  [hep-ph]}}.

\bibitem{Tang:2021lcn}
Y.~Tang and Y.-L. Wu, ``{Conformal transformation with multiple scalar fields
  and geometric property of field space with Einstein-like solutions},''
  \href{http://dx.doi.org/10.1103/PhysRevD.104.064042}{{\em Phys. Rev. D}
  {\bfseries 104} no.~6, (2021) 064042},
  \href{http://arxiv.org/abs/2105.04726}{{\ttfamily arXiv:2105.04726 [gr-qc]}}.

\bibitem{SravanKumar:2018tgk}
K.~Sravan~Kumar and P.~Vargas~Moniz, ``{Conformal GUT inflation, proton
  lifetime and non-thermal leptogenesis},''
  \href{http://dx.doi.org/10.1140/epjc/s10052-019-7449-1}{{\em Eur. Phys. J. C}
  {\bfseries 79} no.~11, (2019) 945},
  \href{http://arxiv.org/abs/1806.09032}{{\ttfamily arXiv:1806.09032
  [hep-ph]}}.

\bibitem{Mikura:2020qhc}
Y.~Mikura, Y.~Tada, and S.~Yokoyama, ``{Conformal inflation in the
  metric-affine geometry},''
  \href{http://dx.doi.org/10.1209/0295-5075/132/39001}{{\em EPL} {\bfseries
  132} no.~3, (2020) 39001}, \href{http://arxiv.org/abs/2008.00628}{{\ttfamily
  arXiv:2008.00628 [hep-th]}}.

\bibitem{Mikura:2021ldx}
Y.~Mikura, Y.~Tada, and S.~Yokoyama, ``{Minimal $k$-inflation in light of the
  conformal metric-affine geometry},''
  \href{http://dx.doi.org/10.1103/PhysRevD.103.L101303}{{\em Phys. Rev. D}
  {\bfseries 103} no.~10, (2021) L101303},
  \href{http://arxiv.org/abs/2103.13045}{{\ttfamily arXiv:2103.13045
  [hep-th]}}.

\bibitem{Banks:1988yz}
T.~Banks and L.~J. Dixon, ``{Constraints on String Vacua with Space-Time
  Supersymmetry},'' \href{http://dx.doi.org/10.1016/0550-3213(88)90523-8}{{\em
  Nucl. Phys. B} {\bfseries 307} (1988) 93--108}.

\bibitem{Banks:2010zn}
T.~Banks and N.~Seiberg, ``{Symmetries and Strings in Field Theory and
  Gravity},'' \href{http://dx.doi.org/10.1103/PhysRevD.83.084019}{{\em Phys.
  Rev. D} {\bfseries 83} (2011) 084019},
  \href{http://arxiv.org/abs/1011.5120}{{\ttfamily arXiv:1011.5120 [hep-th]}}.

\bibitem{Harlow:2018tng}
D.~Harlow and H.~Ooguri, ``{Symmetries in quantum field theory and quantum
  gravity},'' \href{http://dx.doi.org/10.1007/s00220-021-04040-y}{{\em Commun.
  Math. Phys.} {\bfseries 383} no.~3, (2021) 1669--1804},
  \href{http://arxiv.org/abs/1810.05338}{{\ttfamily arXiv:1810.05338
  [hep-th]}}.

\bibitem{Harlow:2018jwu}
D.~Harlow and H.~Ooguri, ``{Constraints on Symmetries from Holography},''
  \href{http://dx.doi.org/10.1103/PhysRevLett.122.191601}{{\em Phys. Rev.
  Lett.} {\bfseries 122} no.~19, (2019) 191601},
  \href{http://arxiv.org/abs/1810.05337}{{\ttfamily arXiv:1810.05337
  [hep-th]}}.

\bibitem{Jimenez:2019woj}
J.~B. Jim\'enez, L.~Heisenberg, and T.~S. Koivisto, ``{The Geometrical Trinity
  of Gravity},'' \href{http://dx.doi.org/10.3390/universe5070173}{{\em
  Universe} {\bfseries 5} no.~7, (2019) 173},
  \href{http://arxiv.org/abs/1903.06830}{{\ttfamily arXiv:1903.06830
  [hep-th]}}.

\bibitem{Armendariz-Picon:1999hyi}
C.~Armendariz-Picon, T.~Damour, and V.~F. Mukhanov, ``{k - inflation},''
  \href{http://dx.doi.org/10.1016/S0370-2693(99)00603-6}{{\em Phys. Lett. B}
  {\bfseries 458} (1999) 209--218},
  \href{http://arxiv.org/abs/hep-th/9904075}{{\ttfamily arXiv:hep-th/9904075}}.

\bibitem{Armendariz-Picon:2000nqq}
C.~Armendariz-Picon, V.~F. Mukhanov, and P.~J. Steinhardt, ``{A Dynamical
  solution to the problem of a small cosmological constant and late time cosmic
  acceleration},'' \href{http://dx.doi.org/10.1103/PhysRevLett.85.4438}{{\em
  Phys. Rev. Lett.} {\bfseries 85} (2000) 4438--4441},
  \href{http://arxiv.org/abs/astro-ph/0004134}{{\ttfamily
  arXiv:astro-ph/0004134}}.

\bibitem{Armendariz-Picon:2000ulo}
C.~Armendariz-Picon, V.~F. Mukhanov, and P.~J. Steinhardt, ``{Essentials of k
  essence},'' \href{http://dx.doi.org/10.1103/PhysRevD.63.103510}{{\em Phys.
  Rev. D} {\bfseries 63} (2001) 103510},
  \href{http://arxiv.org/abs/astro-ph/0006373}{{\ttfamily
  arXiv:astro-ph/0006373}}.

\bibitem{Gialamas:2019nly}
I.~D. Gialamas and A.~B. Lahanas, ``{Reheating in $R^2$ Palatini inflationary
  models},'' \href{http://dx.doi.org/10.1103/PhysRevD.101.084007}{{\em Phys.
  Rev. D} {\bfseries 101} no.~8, (2020) 084007},
  \href{http://arxiv.org/abs/1911.11513}{{\ttfamily arXiv:1911.11513 [gr-qc]}}.

\bibitem{Enckell:2018hmo}
V.-M. Enckell, K.~Enqvist, S.~Rasanen, and L.-P. Wahlman, ``{Inflation with
  $R^2$ term in the Palatini formalism},''
  \href{http://dx.doi.org/10.1088/1475-7516/2019/02/022}{{\em JCAP} {\bfseries
  02} (2019) 022}, \href{http://arxiv.org/abs/1810.05536}{{\ttfamily
  arXiv:1810.05536 [gr-qc]}}.

\bibitem{Garriga:1999vw}
J.~Garriga and V.~F. Mukhanov, ``{Perturbations in k-inflation},''
  \href{http://dx.doi.org/10.1016/S0370-2693(99)00602-4}{{\em Phys. Lett. B}
  {\bfseries 458} (1999) 219--225},
  \href{http://arxiv.org/abs/hep-th/9904176}{{\ttfamily arXiv:hep-th/9904176}}.

\bibitem{Martin:2013uma}
J.~Martin, C.~Ringeval, and V.~Vennin, ``{K-inflationary Power Spectra at
  Second Order},'' \href{http://dx.doi.org/10.1088/1475-7516/2013/06/021}{{\em
  JCAP} {\bfseries 06} (2013) 021},
  \href{http://arxiv.org/abs/1303.2120}{{\ttfamily arXiv:1303.2120
  [astro-ph.CO]}}.

\bibitem{BICEP:2021xfz}
{\bfseries BICEP, Keck} Collaboration, P.~A.~R. Ade {\em et~al.}, ``{Improved
  Constraints on Primordial Gravitational Waves using Planck, WMAP, and
  BICEP/Keck Observations through the 2018 Observing Season},''
  \href{http://dx.doi.org/10.1103/PhysRevLett.127.151301}{{\em Phys. Rev.
  Lett.} {\bfseries 127} no.~15, (2021) 151301},
  \href{http://arxiv.org/abs/2110.00483}{{\ttfamily arXiv:2110.00483
  [astro-ph.CO]}}.

\bibitem{Carlip:2012wa}
S.~Carlip, ``{Challenges for Emergent Gravity},''
  \href{http://dx.doi.org/10.1016/j.shpsb.2012.11.002}{{\em Stud. Hist. Phil.
  Sci. B} {\bfseries 46} (2014) 200--208},
  \href{http://arxiv.org/abs/1207.2504}{{\ttfamily arXiv:1207.2504 [gr-qc]}}.

\bibitem{Witten:2017hdv}
E.~Witten, ``{Symmetry and Emergence},''
  \href{http://dx.doi.org/10.1038/nphys4348}{{\em Nature Phys.} {\bfseries 14}
  no.~2, (2018) 116--119}, \href{http://arxiv.org/abs/1710.01791}{{\ttfamily
  arXiv:1710.01791 [hep-th]}}.

\bibitem{Barcelo:2016xhp}
C.~Barcel\'o, R.~Carballo-Rubio, F.~Di~Filippo, and L.~J. Garay, ``{From
  physical symmetries to emergent gauge symmetries},''
  \href{http://dx.doi.org/10.1007/JHEP10(2016)084}{{\em JHEP} {\bfseries 10}
  (2016) 084}, \href{http://arxiv.org/abs/1608.07473}{{\ttfamily
  arXiv:1608.07473 [gr-qc]}}.

\bibitem{Barcelo:2021idt}
C.~Barcel\'o, R.~Carballo-Rubio, L.~J. Garay, and G.~Garc\'\i{}a-Moreno,
  ``{Emergent gauge symmetries: Yang-Mills theory},''
  \href{http://dx.doi.org/10.1103/PhysRevD.104.025017}{{\em Phys. Rev. D}
  {\bfseries 104} no.~2, (2021) 025017},
  \href{http://arxiv.org/abs/2101.12188}{{\ttfamily arXiv:2101.12188
  [hep-th]}}.

\bibitem{Halliwell:1989myn}
J.~J. Halliwell, ``{INTRODUCTORY LECTURES ON QUANTUM COSMOLOGY},''
\newblock 1989.
\newblock \href{http://arxiv.org/abs/0909.2566}{{\ttfamily arXiv:0909.2566
  [gr-qc]}}.

\bibitem{Hartle:1983ai}
J.~B. Hartle and S.~W. Hawking, ``{Wave Function of the Universe},''
  \href{http://dx.doi.org/10.1103/PhysRevD.28.2960}{{\em Phys. Rev. D}
  {\bfseries 28} (1983) 2960--2975}.

\bibitem{Hawking:1983hj}
S.~W. Hawking, ``{The Quantum State of the Universe},''
  \href{http://dx.doi.org/10.1016/0550-3213(84)90093-2}{{\em Nucl. Phys. B}
  {\bfseries 239} (1984) 257}.

\bibitem{Vilenkin:1982de}
A.~Vilenkin, ``{Creation of Universes from Nothing},''
  \href{http://dx.doi.org/10.1016/0370-2693(82)90866-8}{{\em Phys. Lett. B}
  {\bfseries 117} (1982) 25--28}.

\bibitem{Linde:1983mx}
A.~D. Linde, ``{Quantum Creation of the Inflationary Universe},''
  \href{http://dx.doi.org/10.1007/BF02790571}{{\em Lett. Nuovo Cim.} {\bfseries
  39} (1984) 401--405}.

\bibitem{Vilenkin:1984wp}
A.~Vilenkin, ``{Quantum Creation of Universes},''
  \href{http://dx.doi.org/10.1103/PhysRevD.30.509}{{\em Phys. Rev. D}
  {\bfseries 30} (1984) 509--511}.

\bibitem{Vilenkin:1986cy}
A.~Vilenkin, ``{Boundary Conditions in Quantum Cosmology},''
  \href{http://dx.doi.org/10.1103/PhysRevD.33.3560}{{\em Phys. Rev. D}
  {\bfseries 33} (1986) 3560}.

\bibitem{Vilenkin:1987kf}
A.~Vilenkin, ``{Quantum Cosmology and the Initial State of the Universe},''
  \href{http://dx.doi.org/10.1103/PhysRevD.37.888}{{\em Phys. Rev. D}
  {\bfseries 37} (1988) 888}.

\bibitem{Wang:2019spw}
S.-J. Wang, M.~Yamada, and A.~Vilenkin, ``{Constraints on non-minimal coupling
  from quantum cosmology},''
  \href{http://dx.doi.org/10.1088/1475-7516/2019/08/025}{{\em JCAP} {\bfseries
  08} (2019) 025}, \href{http://arxiv.org/abs/1903.11736}{{\ttfamily
  arXiv:1903.11736 [gr-qc]}}.

\bibitem{Vilenkin:2018dch}
A.~Vilenkin and M.~Yamada, ``{Tunneling wave function of the universe},''
  \href{http://dx.doi.org/10.1103/PhysRevD.98.066003}{{\em Phys. Rev. D}
  {\bfseries 98} no.~6, (2018) 066003},
  \href{http://arxiv.org/abs/1808.02032}{{\ttfamily arXiv:1808.02032 [gr-qc]}}.

\bibitem{Vilenkin:2018oja}
A.~Vilenkin and M.~Yamada, ``{Tunneling wave function of the universe II: the
  backreaction problem},''
  \href{http://dx.doi.org/10.1103/PhysRevD.99.066010}{{\em Phys. Rev. D}
  {\bfseries 99} no.~6, (2019) 066010},
  \href{http://arxiv.org/abs/1812.08084}{{\ttfamily arXiv:1812.08084 [gr-qc]}}.

\end{thebibliography}\endgroup

\end{document}